\documentclass[a4paper,11pt]{article}
\pdfoutput=1 

\usepackage{jcappub}
\usepackage[english]{babel}
\usepackage{amsfonts}
\usepackage{amsmath}
\usepackage{amssymb}
\usepackage{graphicx}
\usepackage{color}
\usepackage{bm}

\def\be{\begin{equation}}
	\def\ee{\end{equation}}
\def\nn{\nonumber\\}
\def\ba{\end{eqnarray}}
\def\ea{\end{eqnarray}}
\def\bea{\begin{eqnarray}}
\def\tea{\end{eqnarray}}

\begin{document}

\title{Primordial Weibel Instability}

\author[a,b,c]{Nahuel Mir\'on-Granese}
\author[c,d]{, Esteban Calzetta}
\author[e,1]{and Alejandra Kandus\note{Corresponding author.}}


\affiliation[a]{Facultad de Ciencias Astronómicas y Geofísicas, Universidad Nacional de La Plata\\ Paseo del Bosque, B1900FWA La Plata, Buenos Aires, Argentina}
\affiliation[b]{Consejo Nacional de Investigaciones Científicas y Técnicas (CONICET)\\Rivadavia 1917, Ciudad de Buenos Aires, Argentina}
\affiliation[c]{Departamento de F\'isica, Facultad de Ciencias Exactas y Naturales, Universidad de Buenos Aires\\
Intendente Güiraldes 2160, Pabellón I Cuidad Universitaria, Ciudad de Buenos Aires (CP 1428), Argentina}
\affiliation[d]{Instituto de Física de Buenos Aires (IFIBA) CONICET-UBA\\ Intendente Güiraldes 2160, Pabellón I Cuidad Universitaria, Ciudad de Buenos Aires (CP 1428), Argentina}
\affiliation[e]{Departamento de Ci\^encias Exatas e Tecnol\'ogicas, Universidade Estadual de Santa Cruz\\ Rodov. Jorge Amado km 16,
CEP: 45.662-900, Ilh\'eus - BA, Brasil}

\emailAdd{nahuelmg@fcaglp.unlp.edu.ar}
\emailAdd{calzetta@df.uba.ar}
\emailAdd{kandus@uesc.br}

\abstract{
We study the onset of vector instabilities in the post-inflationary epoch of the Universe as a mechanism for primordial magnetic fields amplification. 
We assume the presence of a charged spectator scalar field arbitrarily coupled to gravity during Inflation in its vacuum de Sitter state. Gravitational particle creation takes place at the transition from Inflation to the subsequent Reheating stage and thus the vacuum field state becomes an excited many particles one. 
Consequently this state can be described as a real fluid, and we build out the hydrodynamic framework using  second order theories for relativistic fluids with a relaxation time prescription for the collision integral.
Given the high-temperature regime and the vanishing scalar curvature of the Universe during Reheating (radiation-dominated-type era), the fluid can be regarded as a conformal one. 
The large quantum fluctuations induced by the rapid transition from inflationary to effectively radiation dominated expansion become statistical fluctuations whereby both a charge excess and anisotropic pressures are produced in any finite domain. The precise magnitude of the effect for each scale is determined by the size of the averaging domain and the coupling to curvature. We look at domains which are larger than the horizon at the beginning of Reheating, but much smaller than our own horizon, and show that in a finite
fraction of them the anisotropy and charge excess provide suitable conditions for a Weibel instability. If moreover the duration of reheating is shorter than the relaxation time of the fluid, then this instability can compensate or even overcome the conformal dilution of a primordial magnetic field. We show that the non-trivial topology of the magnetic field encoded in its magnetic helicity is also amplified if present.}

\maketitle

\section{Introduction}

Except for the Electroweak (EW) and Quantum Chromodynamic (QCD) phase transitions the epoch of the Universe
prior to  Big-Bang Nucleosynthesis (BBN) is practically unknown \cite{fermilab-rep20,donough-20,CKPWS-18}. Processes such as the Reheating of the Universe after Inflation, which lead to the Radiation epoch are still arena of intense research. Since the seminal works of Albrecht, Steinhardt et al. \cite{AlSteTuWil82}, Kofman, Linde et al
\cite{KoLiSta97} and others, the understanding of the Reheating of the Universe evolved from an almost
instantaneous process to being considered a stage in itself, with several e-folds of duration
\cite{BraNaRa,Loz19,MirtNozAz21}. This fact leads to the question of what processes other than the decay
of the inflaton might take place during that period, specially those related to the evolution of the gravitational field and primordial magnetic fields.

Among different out-of-equilibrium processes in the Early Universe, the onset of instabilities can play a determinant role during Reheating by accelerating the thermalization of the primordial radiation \cite{BraNaRa,Loz19}, 
induce or amplify primordial magnetic fields \cite{calkanma98,FiGru01,KKT11,Ko14,vach20}, and perhaps help 
to trigger phase transitions as the EW one.

In general the research about the first stages of the Universe after Inflation is carried out using non-equilibrium 
quantum field theory tools \cite{CalHu08}, but recent developments in the study of real 
relativistic fluids \cite{Rom-Rom-19} began to pave the road toward the treatment of the matter 
fields in the early Universe within relativistic hydrodynamic formalisms \cite{GraCal,CalKan10}.

Despite meaningful progress in the development of a consistent relativistic hydrodynamics 
\cite{Rom-Rom-19,tanos}, there is no consensus at present on which is the set of appropriate equations to
describe real relativistic fluids. We may recognize two main lines: First Order Theories (FOTs) where
the dynamical variables are the same for ideal and real fluids
\cite{Eck40,LL59,VanBiro12,Nor1,Nor2,Nor3,Kov19,DFNR20,PreRuRe20,HouKov20}
and Second Order Theories (SOTs) where the number of variables for a real fluid is larger than for an ideal one. Typically in SOTs, the dissipative part of the energy momentum tensor is regarded as a hydrodynamic
variable on its own. There are several branches in this line: Israel-Stewart (IS) \cite{Isr76,IS76,IS79a,IS79b,IS80,BiDaHaPubRo20,PaDaBisRoy21a,PaDaBisRoy21b},
Divergence Type Theories (DTTs) \cite{LMR86,GL90,GL91,ReNa95,PRCal09,PRCal10,PRCal13,Cal15,Cal98,LeReRu18,CanCal20,MGKanCal20,Cal20,PeCa21}, 
Denicol-Niemi-Moln\'ar-Rischke (DNMR) theory
\cite{DeKoRi10,BDKMNR11,DNNR11,DMNR12,DNMR12,DNBMXRG14,DN12,MNDR14,ND14}
and Anisotropic Hydrodynamics (AH) \cite{Strick14a,Strick14b,FMRT15,FRST16,NMR17,BHS14,MNR16a,MNR16b}.
For small deviations from equilibrium they all agree and the resulting hydrodynamic formalism is equivalent to taking suitable moments of the relativistic Boltzmann equation with a Grad-like ansatz for the 1-particle distribution function (1pdf) \cite{ChapCow}. 
These theories are quite successful in explaining the results of RHICs. Other scenarios where they can 
certainly be applied are the Early Universe \cite{Nahuel18,nahuel20}
and the Neutron Stars interiors \cite{DexNoNoYuRa21}.

One of the main unsolved cosmological and astrophysical problems is the origin and evolution of magnetic
fields that permeate the Universe \cite{KaKuTs11}.
Magnetogenesis mechanisms that operate at the end of Inflation, as e.g. the one of Refs.
\cite{calkanma98,kacalmawa00} consider that the induced field is subsequently diluted by the cosmological 
expansion. In this manuscript we wonder what processes might act at the same epoch and compensate 
such dilution. Using relativistic hydrodynamics, in this manuscript we want to investigate one such processes,
namely the Weibel instability. 
As a concrete hydrodynamic model, we take the results of Ref. \cite{CalKan16} where 
using a Grad prescription for the 1pdf, the relativistic Weibel instability was naturally 
described.

Weibel instability can be important for magnetogenesis during the primordial QCD phase
transition as was envisaged long time ago, when it was proved (both numerically and analitically) 
that if a cold partonic beam passes through a quark-gluon plasma a beam-driven Weibel instability 
develops and magnetic fields are excited (see section 4.4 of Ref. \cite{WiRyuScle-al12} and references 
therein). The magnetic fields induced by the instability can be quite intense for relatively small
anisotropic velocity distributions, and of coherence length equal to the scale of the patch where
the phase transition occurs.

Since Weibel's seminal work on instabilities due to anisotropies in the velocity distribution function of gases \cite{weibel59, fried1959}, relativistic generalizations were developed in the framework 
of kinetic theory \cite{shlick04,acht07a,acht07b}, and hydrodynamic descriptions for the non-relativistic 
case were worked out not long ago \cite{basu02,BretDeu06,Bret06}.  In addition plasma instabilities related to the quark-gluon plasma were studied in \cite{mrowczynski2006,mrowczynski2017}. In the  model of Ref. \cite{CalKan16}, the anisotropies in the velocity distribution are
encoded in a new tensor fluid variable that represents an anisotropic pressure. 
This SOT hydrodynamic framework has already been used to describe the primordial plasma in the very 
Early Universe and, in particular, to capture its interaction with the primordial gravitational waves \cite{Nahuel18,nahuel20}.

We consider a cosmological scenario where an inflationary period occurred, and during which a charged 
spin-0 field coupled to gravity was in its vacuum state. 
At the transition between Inflation and Reheating, which for simplicity we assume instantaneous, 
the vacuum state becomes an excited, highly populated state \cite{BirrDav82,ParkToms09}.
Due to the high symmetry of the considered stages of the Universe, the quantum gas is also highly
symmetric and consequently all mean values of vector quantities as e.g., momentum flux, 
electric four-current, energy flux, will be vanishing. Quantum fluctuations on the other side, 
build non-vanishing rms deviations. These quantum fluctuations become hydrodynamical fluctuations in the large occupation numbers regime. To obtain the corresponding rms hydrodynamical fluctuations we
apply the Landau prescription, whereby the quantum Hadamard two point functions for the
electric four-current $J^{\mu}$ and spatial components of the energy momentum tensor 
$T^{\mu\nu}$ of the scalar field, averaged on a scale of interest $L_D$, are equated to 
the corresponding two-point correlation functions of a real relativistic plasma. 
The temperature of the fluid is estimated from the vacuum expectation value (VEV) of the scalar 
field energy density $T^{00}$ together with a consistent relation between number and charge densities. 
This estimation and the fact that the scalar curvature of the universe during Reheating is
zero, will indicate whether the fluid can be regarded as conformal or not: if the temperature 
is higher than the effective field mass then we shall consider that it is indeed conformal. 
This entails that phenomena related to the trace of the energy momentum tensor, 
such as bulk viscosity, will be neglected in the present work. 

To be more precise, the diagonal terms in the spatial part of the EMT of the gas of
scalar particles contain information about the isotropic pressure (or hydrodynamic pressure)
as well as about the anisotropic (or non-hydrodynamic) contributions. To estimate the later 
we shall build a Hadamard two-point function of the tensor $T_{tls}^{ij}=T^{ij}-(1/3)g^{ij}T^k_k$, 
i.e., from the spatial components of the EMT without the hydrodynamic components.
After filtering over a scale of physical interest we equate the 
resulting expressions to corresponding fluid variables. Those variables cannot be the usual 
hydrodynamic ones because they do not sustain anisotropic pressures. In SOTs they are readily 
identified as higher moments of the one particle distribution function.

We assume a de Sitter inflationary period after which a Reheating processes took place 
to finally establish the bulk radiation field. According to 
Refs. \cite{Loz19,MirtNozAz21}, the duration of those stages depends on the equation 
of state. The expansion rate is the same as for a radiation dominated universe almost
immediatly after the end of Inflation while the radiation field takes several e-folds to establish.
This means that during that period the charged scalar field and the bulk fields 
can be considered as decoupled, thus allowing the use the one-fluid formalism developed in 
Ref. \cite{CalKan16}.

Whether a Weibel instability will occur on any given domain depends on several occurrences of 
essentially a random nature. The most important conditions are of course enough pressure anisotropy and enough charge excess. The statistics of pressure anisotropy is discussed in Section 
\ref{back-build-a}. It depends on two independent Gaussian parameters whose variance is given in equations 
(\ref{Zast-Z0}) and (\ref{Z02-final}), and on the size of the domain we are looking at in a way given by 
Eq. (\ref{Nijij-lambda}); similarly, charge excess is a third Gaussian parameter whose statistics is 
discussed in Section \ref{back-build}, see Eqs. (\ref{J0-corr-final-b}) and (\ref{J0J0-lambda-b}). 
Fluctuations are larger for small domains and die out for larger ones; of course the Universe as a whole is 
isotropic. For concreteness, we look at domains which are about ten times the size of the horizon at the
beginning of Reheating.
Since there are a huge number of them within our own horizon, in a large number of domains the actual anisotropy
parameters and charge excess will be about or larger than their variances, which we take as an appropriate 
order of magnitude estimate. 

Given a domain where pressure is sufficiently anisotropic, the anisotropic pressure tensor will pick up a preferred
direction, which we may identify with the direction of a particle beam, see the discussion in 
Section \ref{fixingorientations} and Fig. \ref{esquemae+e-}. This direction is essentially equally distributed 
on the unit sphere. The modes which will be most amplified are those which propagate in an orthogonal direction
to the ``beam'', but a perfect alignment is not necessary, see Fig. \ref{svstheta}. This gives enough leeway 
that an instability will likely occur in a significant fraction of the suitable domains. For the actual strength
of the instability see Fig. \ref{sMassMenos}.

The final issue is for how long the instability will be operative. The average pressure anisotropy, which we 
take as a homogeneous background within a given domain, will decay on time scales of the order of the relaxation
time of the fluid. The relaxation time is essentially determined by the fluid temperature, see 
Eq. (\ref{tau-choice}) and Fig. \ref{tau-vs-L}, and the fluid temperature depends in a highly nonlinear way on
the domain size and the coupling to curvature, see Figs. \ref{toofull} and \ref{tempvsL}. On the other hand, 
as reheating progresses the ordinary radiation background in the Universe will build up, and since our fluid 
is charged, it will efficiently thermalize with it. So we need two conditions, the fluid must be close enough 
to free streaming, and moreover the reheating era must last less than the relaxation time. 
We discuss these conditions in Section \ref{weibel-prim}. If these conditions are met, the Weibel instability 
may successfully compensate the dilution of the magnetic field due to the expansion of the Universe, 
see Fig. \ref{results_fig}.

In conclusion, although whether a strong enough Weibel instabily actually happens in a given domain is a random phenomenon, the probabilities are significant enough that we may expect it will actually occur in a large number of domains.

To summarize the results, we found that for different values of the coupling of the scalar field to
gravity the temperature of the fluid is high enough to consider it as conformal during Reheating. 
For the scenarios described in the above paragraphs, we find that for several Reheating models 
\cite{LozAmin16} the instability compensates de dilution due to the expansion at the end of Reheating. 
This sets new initial values of primordial magnetic fiels for their subsequent evolution,

The manuscript is organized as follows. In Section \ref{Fluid-model} we briefly review the derivation of the 
set of tensors and equations for the SOT description of the primordial plasma made in Ref. \cite{CalKan16} and  analyze its conformal invariance. We linearize the equations around an anisotropic background pressure and 
for a generic propagation direction, and obtain a dispersion relation for the vector modes, which now posses 
two sets of solutions associated to two orthogonal polarizations. 
In Section \ref{back-build} we build the simplest matching relations between the fluid and the field two point functions, from which we read the VEVs we need to compute.

In section \ref{scal-phi-sect} we provide the expressions for $T^{\mu\nu}$ and and the four-current $J^{\mu}$. 
(To make the mansucript as self-contained as possible, in Appendix \ref{ChScField} we give a short review of 
charged scalar fields in curved spacetime.) We plot the result of the 
numerical calculation of the VEV $\left\langle T^{00}\right\rangle$,
which shows that the energy density is weakly dependent on the small mass of the scalar field but sensitive 
to the value of the coupling to the gravitational field. We provide the definitions for the noise kernels 
to be used to compute the rms values of the electric charge and anisotropic pressure, and their general
expressions in terms of the field modes as well. 
We compute the filtering on a scale $L_D$ of the resulting expressions of those kernels finding that both 
quantities depend on the averaging scale as $L_D^{-3}$. Regarding the dependence on the coupling to gravity,
we found that the anisotropic pressure depends very weakly on $\xi$, while for the charge excees the dependence
is stronger, going as $\xi^{-1}$. These two are the main parameters that determine the intensity of the instabilites.

In Section \ref{weibel-prim} we apply the results of the previous section to evaluate the fluid
temperature and to study the linear unstable modes from the dispersion relation found in Section 
\ref{Fluid-model}. We confirm that the temperature is high enough to consider the plasma as conformal.
For both polarizations the main instability occurs for propagation orthogonal to the direction where the anisotropic
pressure is maximal. We continue the analysis taking into account only the largest solution to the dispersion relation. As stated in previous paragraphs, we focus on Reheating scenarios whose duration is shorter than the relaxation time of the fluid. The reasons for this assumption are several: (1) The background radiation field to which the scalar plasma would couple is not settled yet, thus we are able to work with a one-fluid formalism; (2) since the  effects of the collisions strongly manifest after the relaxation time $\tau$, we regard an almost collisionless plasma. Under these conditions the Weibel instability is operative during the entire period.

We find that for some models of Reheating, there is a range of values of $\xi$ for which the instability overcomes the dilution by the cosmological expansion at the end of this stage. The associated scales are of the order of the particle horizon at that time. We end this section studying how the topology of the field,
encoded in the magnetic helicity, might be affected by the instability. We find that it is also amplified and that the amplification depends on both unstable frequencies (not only on the largest). This result implies
that the chances of survival of an helical primordial magnetic field are improved.

Section \ref{concl} contains the main conclusions on this work. We work with signature $(-,+,+,+)$ and natural units $\hbar = k_B = c = 1$.

\section{Fluid Description of the Primordial Plasma}\label{Fluid-model}

In this section we shortly review the hydrodynamic description of a conformal plasma developed in 
Ref. \cite{CalKan16}, where we wrote down a minimal SOT to describe a real relativistic magnetized plasma, 
starting from Vlasov-Boltzmann equation and adopting a Grad prescription for the non-equilibrium distribution
function. In this section we particularize the results to a fluid in an expanding FRW space-time and re-obtain
the linear equations for the vector variables considering propagation along an arbitrary direction. There
are two sets of solutions of the dispersion relation that correspond to the two possible polarizations of
the modes and hence two sets of possible unstable momenta.

\subsection{Conformal Plasma in a FRW Background}

Our starting point is the relativistic Vlasov-Boltzmann equation
\begin{equation}
p^{\mu}\left[\nabla_\mu - e F_{\mu\rho}\frac{\partial}{\partial p_{\rho}}\right]f\left(x^{\mu},p^{\mu}\right)
= I_{col} \left(x^{\mu},p^{\mu}\right) \label{II-a}
\end{equation}
for a one particle distribution function $f\left(x^{\mu},p^{\mu}\right)$.
As we are interested in small deviations from equilibrium, we write
$f$ in the form of a Grad-like ansatz, i.e.,
\begin{equation}
f\left(x^{\mu},p^{\mu}\right) = f_{eq}\left[ 1 + Z \right] \label{II-b}
\end{equation}
where $f_{eq}$ is the Maxwell-J\"utner distribution function, and where
\begin{equation}
Z = \frac{\tau}{2\left\vert u_{\rho}p^{\rho}\right\vert}\left[\zeta_{\rho\sigma}p^{\rho}p^{\sigma} + e\zeta_{\rho}p^{\rho}
+ \frac{2}{4T}\tanh\left(\alpha\right) ~ \zeta_{\rho}u_{\sigma} p^{\rho}p^{\sigma}\right] \label{II-c}
\end{equation}
accounts for deviations from equilibrium \cite{CalKan16}.
The tensors $\zeta_{\rho}$ and $\zeta_{\rho\sigma}$ are the new variables
that encode the non-ideal features of the plasma: the former represents
conduction currents and the latter the viscous stresses. To properly describe a
conformal fluid they must satisfy the constraints: $\zeta^{\mu}_{\mu} =
\zeta^{\mu\nu}u_{\nu} = \zeta^{\mu}u_{\mu}=0$.
The quantity $\alpha$ is the thermal potential and $\tau$ a relaxation time. By taking moments of the 
distribution function we build the fundamental tensors of the model whose corresponding evolution equations 
are obtained by taking suitable averages of the Vlasov-Boltzmann equation. The resulting theory may be regarded as a linearized approximation within a more general SOT scheme, such as DTTs \cite{LMR86,GL90,GL91,ReNa95,PRCal09,PRCal10,PRCal13,Cal15,Cal98,LeReRu18,CanCal20,MGKanCal20}, DNMR theory \cite{DeKoRi10,BDKMNR11,DNNR11,DMNR12,DNMR12,DNBMXRG14,DN12,MNDR14,ND14}
or anisotropic hydrodynamics \cite{Strick14a,Strick14b,FMRT15,FRST16,NMR17,BHS14,MNR16a,MNR16b}. 

The general form of the tensors that describe the magneto-hydrodynamics is
\begin{equation}
A^{\mu_1\cdots\mu_n}_s = \int Dp\left(\mathrm{sign}\left[p^0\right]\right)^s p^{\mu_1} \cdots p^{\mu_n}f
\end{equation}
with $s=0,1$ and the general conservation law reads
\begin{equation}
A^{\mu\mu_1\cdots\mu_n}_{s;\mu} - e\sum_{i=1}^n F^{\mu_i}_{\mu}
A_{s}^{\mu\mu_1\cdots (\mu_i)\cdots\mu_n} = -I_s^{\mu_1\cdots\mu_n}\,,
\end{equation}
where the notation $(\mu_i)$ means that $\mu_i$ is excluded, the tensor $I_s^{\mu_1\cdots\mu_n}$
represents dissipation and it is given by
\begin{equation}
I_s^{\mu_1\cdots\mu_n} = -\int Dp\left(\mathrm{sign}\left[p^0\right]\right)^s
p^{\mu_1} \cdots p^{\mu_n} I_{col}\,,
\end{equation}
with the invariant phase-space measure
\bea
Dp=\frac{d^4p}{(2\pi)^3}\delta(p^2)=\frac{d^3p}{(2\pi)^3p^0}\left[\delta(p^0-p)+\delta(p^0+p)\right]\,.
\tea

The fundamental tensors we work with in this paper are the charge current $J^{\mu}$, 
the energy momentum tensor $T^{\mu\nu}$, two non-conserved tensors, $A^{\mu\nu}$ and $A^{\mu\nu\sigma}$, whose divergencies provide the equations for  Ohmic dissipation and
viscous stresses respectively, and the Maxwell tensor $F_{\mu\nu}$. The number current $N^{\mu}$ is automatically conserved in a conformal theory. 

We shall consider the evolution of the real relativistic plasma in a spatially flat Friedman-Robertson-Walker 
(FRW) Universe. 
In conformal coordinates $x^\mu=(\eta,\vec x)$, the metric tensor is $g_{\mu\nu} = a^2(\eta)\tilde g_{\mu\nu}$. It is straightforward to show that the hydrodynamic theory under study is conformally invariant. The conformal transformations that apply in this case are
$T=\tilde T/a(\eta)$, $\alpha=\tilde\alpha$, $u^{\mu} = \tilde u^{\mu} /a(\eta)$,
$\tau = a(\eta)\tilde \tau$, $\zeta^{\mu} =
a^{-2}(\eta)\tilde\zeta^{\mu}$, $\zeta^{\mu\nu} = a^{-2}(\eta)\tilde\zeta^{\mu\nu}$ and 
$F^{\mu\nu} = a^{-4}(\eta)\tilde F^{\mu\nu}$. Moreover, to write down dimensionless equations we choose the Hubble constant during Inflation ($H$) as the fiducial dimensionful energy scale. The dimensions of each variables are $[\tilde T]={\rm energy}$, $ [\tilde u^{\mu}] = 1$ , $ 	[\tilde\tau ] = {\rm energy}^{-1} $, 
$[\tilde\zeta^{\mu}] ={\rm energy}$, $[\tilde\zeta^{\mu\nu}] = 1$, 
$[\tilde F^{\mu\nu}] = {\rm energy}^2$. So we define the dimensionless quantities $\Upsilon=\tilde T / H$, $\hat\tau = \tilde \tau H$,
$Z^\mu=\tilde\zeta^\mu /H  $, $ Z^{\mu\nu}=\tilde \zeta^{\mu\nu}$ and $ \hat F^{\mu\nu}=\tilde F^{\mu\nu} /H^2$. For the coordinates we apply the transformation $x^{\mu} \to  x^{\mu} /H$, so each derivative (either time or spatial) will be accompanied by a factor $H$.

Explicitly we have
\begin{equation}
\hat A^{\mu}_0 = \hat J^{\mu} = \Upsilon^3\left(e q_1 \tilde u^{\mu} + \Lambda e^2 \hat\tau Z^{\mu} \right)\label{II-d1-a}
\end{equation}
\begin{equation}
\hat A^{\mu}_1 = \hat N^{\mu} = \Upsilon^3 \left(q_2 \tilde u^{\mu} + e\hat \tau\Lambda Z^{\mu}\right)
\label{II-d1N}
\end{equation}
\begin{equation}
\hat A^{\mu\nu}_0 = \hat T^{\mu\nu} = 3q_2 \Upsilon^4\left[\tilde u^{\mu}\tilde u^{\nu}
+ \frac{1}{3}\tilde h^{\mu\nu}\right] + \frac{4q_2}{5}\hat\tau \Upsilon^5 Z^{\mu\nu}\label{II-d2-a}
\end{equation}
\begin{equation}
\hat A^{\mu\nu}_1 \equiv \hat A^{\mu\nu}  = 3q_1 \Upsilon^4\left[\tilde u^{\mu}\tilde u^{\nu} 
+ \frac{1}{3}\tilde h^{\mu\nu}\right] 
+ \kappa_1 e \hat\tau \Upsilon^4 \left[Z^{\mu} \tilde u^{\nu} + Z^{\nu} \tilde u^{\mu}\right]
+ \eta_1 \hat\tau \Upsilon^5 Z^{\mu\nu}\label{II-d3}
\end{equation}
and
\begin{eqnarray}
\hat A^{\mu\nu\sigma}_1\equiv\hat A^{\mu\nu\sigma} &=& 12 q_2\Upsilon^5\left[ \tilde u^{\mu}\tilde u^{\nu}\tilde u^{\sigma}
+ \frac{1}{3}\left(\tilde h^{\mu\nu}\tilde u^{\sigma}
+ \tilde h^{\mu\sigma}\tilde u^{\nu} + \tilde h^{\nu\sigma}\tilde u^{\mu}\right)\right]\nn
&-& \frac{q_1}{2}e\hat\tau \Upsilon^5\left[Z^{\mu}\tilde u^{\nu}\tilde u^{\sigma} + Z^{\nu}\tilde u^{\sigma}\tilde u^{\mu}
+ Z^{\sigma} \tilde u^{\mu} \tilde u^{\nu}
+ \frac{1}{5}\left(\tilde h^{\mu\nu}Z^{\sigma} + \tilde h^{\mu\sigma}Z^{\nu}
+ \tilde h^{\nu\sigma}Z^{\mu}\right)\right]\nn
&+& 4\hat\tau q_2 \Upsilon^6\left(Z^{\mu\nu}\tilde u^{\sigma} + Z^{\mu\sigma}\tilde u^{\nu}
+ Z^{\nu\sigma}\tilde u^{\mu}\right) \label{II-d4}
\end{eqnarray}
where
\bea
q_1&=& (2/\pi^2)\sinh\tilde\alpha\,,\label{q1}\\
q_2 &=& (2/\pi^2)\cosh\tilde\alpha\,,\label{q2}
\tea
$\eta_1 = 4q_1/5$,
$\kappa_1 = 2/(\pi^4 q_2)$ and $\Lambda = (4q_2^2 - 3q_1^2)/24q_2$. Clearly exprs. (\ref{q1}) and 
(\ref{q2}) satisfy the constraint 
\begin{equation}
q_2^2 - q_1^2 = \frac{4}{\pi^4}\, , \label{const-q1q2}
\end{equation}
which will be used in Section \ref{weibel-prim} to evaluate the plasma temperature.
The electromagnetic tensor is
\begin{equation}
\hat F_{\mu\nu} =  \hat E_{\nu} \tilde u_{\mu} - \hat E_{\mu} \tilde u_{\nu}
+ \tilde \eta_{\mu\nu\rho\sigma} \hat B_{\rho} \tilde u^{\sigma} \label{II-d5}\,,
\end{equation}
where $\tilde \eta_{\mu\nu\rho\sigma}$ is the four-dimensional totally antisymmetric tensor. 
The moments of the collision integral that appear in the r.h.s. of the 
conservation equations are \cite{CalKan16}
\begin{eqnarray}
I_0 &=& I_1 = I^{\mu}_0 = 0 \label{I-0}\\
\hat I^{\mu}_1 = \hat I^{\mu} &=& e \kappa_1 \Upsilon^4 Z^{\mu}\label{I-1}\\
\hat I^{\nu\rho}_1 = \hat I^{\nu\rho} &=& -\frac{q_1}{2} e \Upsilon^5 \left(Z^{\nu} \tilde u^{\rho}
+ Z^{\rho}\tilde u^{\nu}\right) + 4q_2 \Upsilon^6 Z^{\nu\rho}\label{I-2}
\end{eqnarray}

The conservation equations of the theory are
\begin{eqnarray}
\hat J^{\mu}_{;\mu} &=& 0 \label{II-f1} \\
\hat T^{\mu\nu}_{;\nu} - \hat F^{\mu\nu} \hat J_{\nu} &=& 0 \label{II-f2} \\
\tilde h^\mu{}_\alpha \left[\hat A^{\alpha\nu}_{;\nu} - e \hat F^{\alpha\nu} \hat N_{\nu} \right]
&=& - \tilde h^\mu{}_\alpha \hat I^{\alpha \nu} \label{II-f3} \\
\tilde \Lambda^{\mu\nu}{}_{\alpha\beta}\,\left[\hat A^{\alpha\beta\sigma}_{;\sigma} - e \hat F^{\alpha\sigma} 
{\hat A^{\beta}}_{\sigma} - e \hat F^{\beta\sigma} {\hat A^{\alpha}}_{\sigma} \right]&=& - \tilde \Lambda^{\mu\nu}{}_{\alpha\beta}\,\hat I^{\alpha\beta} \label{II-f4}\\
\hat F^{\mu\nu}{}_{;\nu} &=& 4\pi \hat J^{\mu} \label{II-f6}\,\\
\tilde \eta^{\alpha\mu\nu\rho} \left[ \hat F_{\mu\nu;\rho} + \hat F_{\nu\rho;\mu} 
+ \hat F_{\rho\mu;\nu}\right] &=& 0 \label{II-f5}\,,
\end{eqnarray}
where
\bea
\tilde h_{\mu\nu}=\tilde g_{\mu\nu}+\tilde u_\mu \tilde u_\nu
\tea
is the transverse projector and
\bea
\tilde \Lambda^{\mu\nu}{}_{\alpha\beta}=\frac12\left[ \tilde h^\mu{}_\alpha \tilde h^\nu{}_\beta
+ \tilde h^\mu{}_\beta \tilde h^\nu{}_\alpha-\frac23 \tilde h^{\mu\nu}\tilde h_{\alpha\beta}\right]
\tea
is the symmetric, double transverse and traceless projector. The non-equilibrium equations (\ref{II-f3}) and
(\ref{II-f4}) must be projected in order to provide the right number of supplementary equations \cite{CalKan16}.

\subsection{Linear Analysis}

To analyze the evolution of linear perturbations we propose the following
decomposition:
\begin{eqnarray}
\tilde g_{\mu\nu} &=& \eta_{\mu\nu}\label{metricaconforme}\\
\tilde u^{\mu} &=&  U_0^{\mu} + v^{\mu} \label{u-lin}\\
Z^{\mu} &=& 0 + z^{\mu} \label{zmu-lin}\\
Z^{\mu\nu} &=& Z_0^{\mu\nu} + z^{\mu\nu}  \label{zmunu-lin}\\
\hat F^{\mu\nu} &=& 0 + f^{\mu\nu} \label{emt-lin}\\
\tilde \alpha &=& \alpha_0 + \alpha_1 \label{alp-lin}\\
\Upsilon &=& \Upsilon_0 + t \label{temp-lin}
\end{eqnarray}
where $\eta_{\mu\nu}$ is the unperturbed Minkowski metric and $U_0^{\mu}$ is the four-velocity of the fiducial observers from which we define the electric and magnetic fields. Furthermore $U_0^\mu$ defines the time direction and the time derivative through the partial derivative along
$U_0^{\mu}$ as $U_0^{\mu}X^{\beta}_{,\mu} = dX^{\beta}/d\eta$. We also consider $U_0^{\mu}v_{\mu}=0$ and
$U_0^{\mu}U_{0\mu}=-1$, and construct the zeroth order spatial projector $\Delta^{\mu\nu}=U_0^{\mu}U_0^\nu+\eta^{\mu\nu}$. This four-velocity for comoving observers is $U_0^{\mu}=(1,0,0,0)$. We use $U_0^\mu$ and 
$\Delta^{\mu\nu}$ in order to split the system into its temporal and spatial projections.

Given the relations (\ref{metricaconforme}) and (\ref{u-lin}) we expand $\tilde h^{\mu\nu} =\Delta^{\mu\nu}
+ U_0^{\mu}v^{\nu} + U_0^{\nu}v^{\mu} + \mathcal{O}(2)$. For the Maxwell tensor we write
\begin{eqnarray}
f^{\mu\nu} &=& U_0^{\mu} \epsilon^{\nu}
- U_0^{\nu} \epsilon^{\mu}
+\eta^{\mu\nu}_{~~\sigma\rho} b^{\sigma} U_0^{\rho}\,.
\label{fmunu1stor}
\end{eqnarray}

\subsubsection{Solving the Vector Sector}

To solve eqs (\ref{II-f1})-(\ref{II-f5}) we propose a solution that varies as
$\exp\left(\sigma \eta+i\bar k.\bar r\right)$, we extract the vector sector of the theory
from the equations for the momentum, conduction current and Amp\`ere's law
and the equation for the viscous stress tensor. We only consider the terms which contain vector degrees of freedom, disregarding the first order scalar quantities, in order to capture the simplest dynamical behavior of the vector sector. We first write down the equations in a coordinate system in which the background anisotropic pressure tensor reads
$Z_0^{ij} = \mathrm{diag}(Z_0+Z_{*},Z_0-Z_{*},-2Z_0)$ while the spatial propagation direction $\vec k$ remains arbitrary. From eqs. (\ref{II-d2-a}) and (\ref{zmunu-lin}) we see that $Z_0$ and $Z_{*}$ must satisfy  
$-5/4 \leq \hat\tau \left(Z_0\pm Z_{*} \right)$ and $\hat\tau Z_0 \leq 5/8$ for the pressure to be 
non-negative . The notation $Z_0\pm Z_{*}$ simultaneously encodes two quantities $Z_0+ Z_{*}$ and $Z_0- Z_{*}$ and the inequality must be fulfilled for both. We adopt the more stringent bound $-5/4 \leq \hat\tau \left(Z_0\pm Z_{*} \right)\leq 5/8$. Defining $Q=q_{10}/q_{20}$ the set of vector equations now reads
\begin{equation}
0 = \Upsilon_0\sigma \eta_{ijl}\hat k_jv_l(k) + \frac{1}{5} i \Upsilon_{0}^2
\hat\tau k \eta_{ijl} \hat k_j \hat k_m z^{m}_l(k) + \frac{1}{4}e Q 
\eta_{ijl}\hat k_j f_{0l}(k) \label{vec-eq1}
\end{equation}
\begin{eqnarray}
0 &=& 4 Q \Upsilon_0 \sigma \eta_{ijl} \hat k_j v_l(k)
+ \frac{1}{2}e\Upsilon_0\left(1-Q^2\right)\left(1+\hat\tau\sigma\right)\eta_{ijl}
\hat k_jz_l(k)\nonumber\\
&+&\frac{4}{5} i\hat\tau Q\Upsilon_0^2k\eta_{ijl}\hat k_j \hat k_m z^m_l(k)+ e\eta_{ijl} \hat k_j f_{0l}(k) \label{vec-eq2}
\end{eqnarray}
\begin{eqnarray}
0 &=& 4q_{20} \Upsilon_0  \eta_{iqp} \hat k_q  \hat k_j z_{pj}(k) 
+ 4 q_{20} i k  \eta_{iqp} \hat k_q v_{p}(k)
- \frac{e \hat\tau}{10} q_{10} ik \eta_{iqp} \hat k_q  z_{p}(k)
\nonumber\\
&+& 4 q_{20}\Upsilon_0 \hat\tau \left[ ik   \hat k_j Z_{0j}^l \hat k_l \eta_{iqp} \hat k_q v_{p}(k) 
+ \sigma \eta_{iqp} \hat k_q  \hat k_j  z_{pj}(k) \right]\nonumber\\
&+& \frac{2}{5} e q_{10}  \hat\tau 
\left[2\hat k_q \hat k_p Z_{0qp} \eta_{ijl}f_{jl} +  \left(Z_{0ip} - \hat k_i \hat k_q Z_{0pq} \right) \eta_{pjl}f_{jl} \right]
\label{vec-eq3-c}\,.
\end{eqnarray}
\begin{equation}
0 = \sigma \eta_{kli} \hat k_l f_{0i}(k) 
-  \frac{i k}{2} \eta_{klj}f_{lj}(k) 
- 4\pi \Upsilon_0^3 \left[ e q_{10} \eta_{kli} \hat k_l v_i(k)
+ \Lambda_0 e^2 \hat\tau \eta_{kli} \hat k_l z_i (k)\right] \label{eq-vec4}\,,
\end{equation}
and
\begin{equation}
\frac{\sigma}{2} \eta_{ilj}f_{lj}(k) - i k \eta_{ilj} \hat k_l f_{0j}(k)=0 \label{eq-vec5}\,.
\end{equation}

The expression between rounded brackets in the last line of eq. (\ref{vec-eq3-c}) still couples the different components of the magnetic field but it is easier to deal with. We analyze this issue and the decoupling of the different components of the vector quantities in the following subsection.

\subsection{Fixing the propagation and background orientations}\label{fixingorientations}

The system of equations (\ref{vec-eq1}), (\ref{vec-eq2}), (\ref{eq-vec4}), (\ref{eq-vec5}) and (\ref{vec-eq3-c}) which describes the linear dynamics of the vector sector is written using a coordinate system in which  
$Z_{0ij} =  {\rm diag} \left(Z_0+Z_*, Z_0-Z_*,-2Z_0\right)$ and, without loss of generality $Z_0,~Z_* > 0$ and $Z_* < 3Z_0$, 
and the propagation vector $\vec k$ remains arbitrary. Since eq. (\ref{vec-eq3-c}) mixes and couples the different spatial components of the magnetic field, here we show how to decouple them. 
We begin by setting a rotation transformation 
${\cal O}$ to rotate and fix the propagation direction such that
$\vec k\to\vec k' = \mathcal{O} \,\vec k$ with $\vec k' = (0,0,k)$. 
Let $\theta$ and $\varphi$ be the usual polar and azimuthal angles which describe the arbitrary propagation direction $\hat k =(\cos\varphi\sin\theta,\sin\varphi\sin\theta,\cos\theta)$. In fact we have to transform all the other vectors and tensors, for instance the anisotropic pressure tensor transforms as $Z_{0ij}\to Z'_{0ij} = {\cal O}_{ik} Z_{0kl} {\cal O}^{-1}{}_{lj}$. Hereafter a 'prime' denotes a transformed quantity. We are only interested in the components of the dynamical variables perpendicular to the propagation direction. Since for the new rotated quantites $\hat k'=\hat z'$, we consider just the 
$x'-y'$ directions.
It turns out that the tensor which mixes the directions in eq. (\ref{vec-eq3-c}) now reads $Z'_{0ij} - \hat k'_i\hat k'_l Z'_{0lj}$. This is not a symmetric tensor but it is straightforward to see that the $x'-y'$ block does satisfy the symmetry condition, so we can diagonalize it to obtain the eigenvalues
\begin{eqnarray}
\lambda_{\pm} &=& \frac{1}{2}\left[Z_0\left(3\cos^2\theta - 1\right)-Z_*\sin^2\theta\cos(2\varphi)\right]
\pm \Bigg\{\frac{9}{4} Z_0^2 \sin^4\theta \nonumber\\
&-& \frac{3}{2} Z_0Z_* \left(1+\cos^2\theta\right)\sin^2\theta\cos(2\varphi)
+ Z_*^2 \left[\frac{1}{4} \sin^4\theta\cos^2(2\varphi) + \cos^2\theta\right]\Bigg\}^{1/2}
\label{eigenvalue}
\end{eqnarray}
with the corresponding normalized eigenvectors
\begin{equation}
\bar \varepsilon_{\pm} = 
\frac{\mp Z_*\left\vert\cos\theta\sin(2\varphi)\right\vert}{\sqrt{ \left[Z_0-Z_*\cos(2\varphi)-\lambda_{\pm}\right]^2
		+ \left[Z_*\cos\theta\sin(2\varphi)\right]^2 }}
\left(\frac{Z_0-Z_*\cos(2\varphi)-\lambda_{\pm}}{Z_*\cos\theta\sin(2\varphi)};1;0\right).
\label{eigenvector}
\end{equation}

For the complete tensor $Z'_{0ij} - \hat k'_i\hat k'_l Z'_{0lj}$ there is a third eigenvalue $\lambda_0=0$ with its corresponding left eigenvector $\hat k'$. We can build a new rotated orthonormal basis made of $\{\bar\varepsilon_{+}, \bar\varepsilon_{-}, \hat k'\}$. Therefore considering only the components of the vector dynamical variables which belong 
to the $x'-y'$ plane, the mixing term in eq. (\ref{vec-eq3-c}) can be rewritten in the new basis as a diagonal tensor, 
namely
\bea
\left(Z'_{0ip} - \hat k'_i \hat k'_q Z'_{0pq} \right) \eta'_{pjl}f'_{jl}= \lambda_{(i)} \eta'_{ijl}f'_{jl}(k)
\label{lastterm-2}
\tea
where $\lambda_{(i)}$ denotes the eigenvalues $\lambda_\pm$ corresponding to the directions $\bar\varepsilon_\pm$ respectively. The brackets around the subindex $(i)$ indicate that there is no sum over $i$. The eq. (\ref{vec-eq3-c}) for the $\bar\varepsilon_\pm$ directions then becomes
\begin{eqnarray}
0 &=& 4q_{20} \Upsilon_0  \eta'_{iqp} \hat k'_q  \hat k'_j z'_{pj}(k) +4 q_{20}\Upsilon_0 \hat\tau \left[ik   \hat k'_j Z'_{0j}{}^l \hat k'_l \eta'_{iqp} \hat k'_q v'_{p}(k)
+ \sigma \eta'_{iqp} \hat k'_q  \hat k'_j  z'_{pj}(k) \right]\nonumber\\
&-& \frac{e \hat\tau}{10} q_{10} ik \eta'_{iqp} \hat k'_q  z'_{p}(k) +  4 q_{20} i k  \eta'_{iqp} \hat k'_q v'_{p}(k)+\frac{2}{5} e q_{10}  \hat\tau 
\left[2\hat k'_p \hat k'_q Z'_{0pq} + \lambda_{(i)}\right] \eta'_{ijl}f'_{jl}(k)
\label{vec-eq3}.
\end{eqnarray}

Finally, the system of equations for the vector dynamical variables is composed by (\ref{vec-eq1}), (\ref{vec-eq2}), (\ref{eq-vec4}), (\ref{eq-vec5}) and (\ref{vec-eq3}). These relations must be understood 
as equations for the rotated (primed) quantities in the directions $\bar \varepsilon_{+}$ and 
$\bar \varepsilon_{-}$ of the new basis.

\begin{figure}
	\centering
	\includegraphics[width=8cm]
	{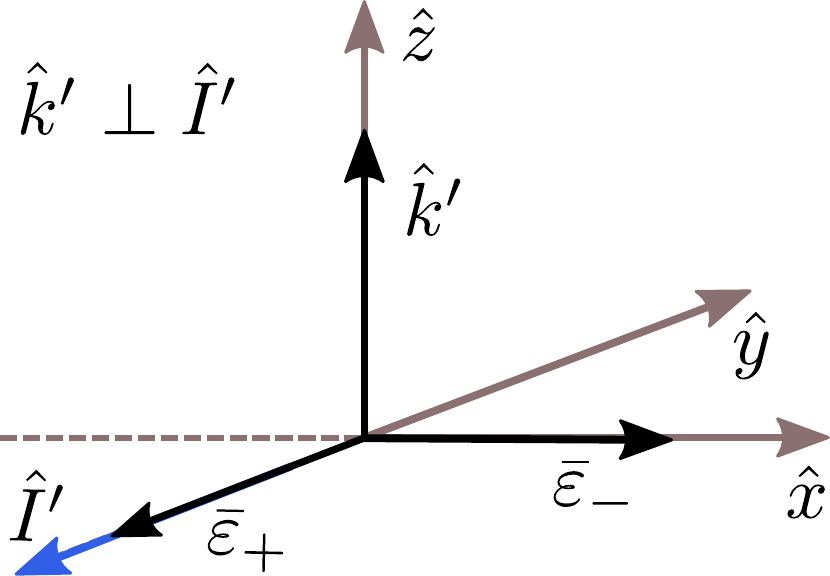}
	\caption{Graphical representation of the propagation direction $\hat k'$ and the transverse components $\bar\varepsilon_+$ and $\bar\varepsilon_-$ of the magnetic field in an extreme case where we consider a particle beam propagating in the direction $\hat I'$ perpendicular to $\hat k'=\hat z$. To accomplish this condition we fix the orientation of $Z'_{0ij}$ (and thus that of $\hat I'$) by fixing the angles defined in Section \ref{fixingorientations} such that $\varphi=\pi/2$ and $0\leq\theta\leq\pi$. We observe that $\bar \varepsilon_{-}=\hat k'\times\hat I'=\hat x$ and $\bar \varepsilon_{+}=\hat I'=-\hat y$ for the entire range of $\theta$.}
	\label{esquemae+e-}
\end{figure}

In figure \ref{esquemae+e-} we show a graphical representation of the vectors $\hat k'$, $\bar\varepsilon_+$ and $\bar\varepsilon_+$ regarding a concrete example as we explain in the next paragraphs.

Since we define $Z_{0ij}={\rm diag}(Z_0+Z_*,Z_0-Z_*,-2Z_0)$ with $Z_0,Z_*>0$ and $Z_*<3Z_0$, the expressions for the distribution function (\ref{II-b}) and (\ref{II-c}) determine that most of the particles propagate through the direction $\hat x$. Considering an extreme case we could have a propagating beam in the direction $\hat I$ with $\hat I \parallel \hat x $. We express all our results for rotated (primed) quantities, in this way the particle beam propagates through the direction $\hat I'=O \hat I=O \hat x$, where $O$ is the matrix rotation defined at the beginning of Section \ref{fixingorientations}, and the magnetic field propagates in $\hat k' = O \hat k=\hat z$.

To be concrete let us consider that the propagation direction of the magnetic field $\hat k'$ is perpendicular to $\hat I'$, for instance we take $\varphi=\pi/2$ and $0\leq\theta\leq\pi$. In this case $\hat k'=\hat z$ and $\hat I'=-\hat y$ for the entire range of $\theta$. Moreover the largest unstable component of the magnetic field $\bar \varepsilon_{-}$ (see Section \ref{dispersionrelation} and Figure \ref{sMassMenos}) corresponds to $\bar \varepsilon_{-}=\hat k'\times\hat I'=\hat x$, while the component $\bar \varepsilon_{+}=\hat I'=-\hat y$ as we observe in Figure \ref{esquemae+e-}. Furthermore in this scheme, the largest unstable growing component of the magnetic field $\bar\varepsilon_-$ coincides with the growing component of the magnetic field related to transverse plasma waves reported in \cite{fried1959}. 

\subsection{Dispersion Relation}\label{dispersionrelation}

In this section we compute the dispersion relation of the system made up of eqs. (\ref{vec-eq1}), 
(\ref{vec-eq2}), (\ref{eq-vec4}), (\ref{eq-vec5}) and (\ref{vec-eq3}).
We consider the dynamical equations for the vector components related to the directions of the eigenvectors 
$\bar \varepsilon_{\pm}$. All the vector and tensor quantities correspond to the rotated (primed) ones.

We define $s=\sigma\hat\tau$, 
$\kappa = k' \hat\tau$, $\varpi = \pi e^2 \hat\tau^2 q_{20} \Upsilon_0^2$ and then the sought relation reads
\begin{eqnarray}
0 &=& s_{\pm}^5 + 2 s_{\pm}^4 + \left[1 + \frac{\kappa^2}{5}\left(6 + Z'_{0zz}
\hat\tau\Upsilon_0\right)
+ \frac{4 \varpi}{3}  \right] s_{\pm}^3
+ \left[\frac{\kappa^2}{5}\left(11 + Z'_{0zz} \hat\tau \Upsilon_0\right) 
+ \varpi Q^2 + \frac{4\varpi}{3} \right] s_{\pm}^2 \nonumber\\
&+& \Bigg[\frac4{25} \varpi\kappa^2\left(\frac53\left(1+\hat \tau\Upsilon_0Z'_{0zz}\right)-Q^2\left(1-\hat\tau\Upsilon_0\left(\frac34 Z'_{0zz}+\lambda_\pm\right)\right)\right)+ \kappa^2  +
\varpi Q^2\nonumber\\
&&+ \frac{\kappa^4}{5}\left(1 + Z'_{0zz}\hat\tau\Upsilon_0\right) \Bigg] s_{\pm} +\frac{\kappa^4}{5}\left(1 + Z'_{0zz}\hat\tau \Upsilon_0\right)
+ \frac{4}{25}\varpi Q^2\hat\tau\Upsilon_0\kappa^2\left(2 Z'_{0zz}+\lambda_{\pm}\right)
\label{disp-rel-pm}
\end{eqnarray}
where subindex $+(-)$ corresponds to the dynamics of the vector components along the direction of the eigenvector 
$\bar\varepsilon_+ (\bar\varepsilon_-)$. Observe that we have two sets of solutions, each corresponding to
either $+$ or $-$. Instabilities emerge as long as there exists a positive real solution to the equation (\ref{disp-rel-pm}) and/or a complex solution with positive real part. 
In Section \ref{weibel-prim} we discuss the explicit positive roots of (\ref{disp-rel-pm}) with fixed parameters corresponding to the cosmological scenario to be described in Section \ref{scal-phi-sect}.

So far we set the magneto-hydrodynamic framework to study the primordial Weibel instability. We must now
obtain the values of the different parameters of the fluid from a concrete particle physics model.

\section{Building the Background}\label{back-build}

We want to study the evolution of vector magneto-hydrodynamic perturbations of a scalar plasma
during Reheating, i.e. the epoch between the end of Inflation and the establishment of the bulk 
radiation field. 

Let us consider a domain of size $L_D$ where the state of the fluid may be regarded
as mostly homogeneous, though anisotropic. The values of the hydrodynamic variables will
be different for each domain and therefore we can only regard them as stochastic. 
Our line of work is to find the first and second moments of these variables and 
then assume that we are looking at a typical domain.
From Section \ref{Fluid-model} we see that the fluid variables are the thermal
potential $\alpha$, the temperature $T$, the four-velocity $u^{\mu}$ and the 
dissipative currents $Z^{\mu}$ and $Z^{\mu\nu}$. 
For the fluid under consideration, it is natural to assume that $\left\langle Z^{\mu}\right\rangle = \left\langle
Z^{\mu\nu}\right\rangle = 0$. For the other variables we have $\alpha = \left\langle\alpha\right\rangle 
+ \alpha_{(s)}$, $T = \left\langle T \right\rangle + T_{(s)}$ and  $u^{\mu} = \left\langle u^{\mu} \right\rangle 
+ u^{\mu}_{(s)}$, with $\left\langle \tilde u^{\mu} \right\rangle = \left\langle \tilde u^{0} \right\rangle U^{\mu}$,
where $U^{\mu} = \left(1,\vec 0\right)$. Using the constraints $\hat u^2 = -1$ and $\hat u_{\mu}Z^{\mu} = 
\hat u_{\mu}Z^{\mu\nu} = Z^{\mu}_{\mu} = 0$ we obtain
\begin{eqnarray}
\left\langle \tilde u^{0} \right\rangle^2 &=& 1 
- \left\langle \left(u^{0}_{(s)}\right)^2\right\rangle 
+ \left\langle u^{i}_{(s)} u_{(s) i}\right\rangle \label{III-1} \\
0 &=&  \left\langle u_{(s)\mu} Z^{\mu}\right\rangle
= \left\langle u_{(s)\mu} Z^{\mu\nu}\right\rangle \label{III-2}
\end{eqnarray}
The dimensionless conserved conformal currents are eqs. (\ref{II-d1-a}) and
(\ref{II-d2-a}).

It is clear that the quantum state of the field is rotationally invariant 
and so will be the expectation values derived from it. Moreover, the initial
state is charge-neutral, and the gravitational particle creation process
does not pick a preferred charge sign. Consequently the expectation values
must be even under charge reversal. Let us call $\hat J^{\mu}_{(Q\lambda)}$
and $\hat T^{\mu\nu}_{(Q\lambda)}$ the Heisenberg current and the EMT
averaged over a scale $\lambda$. We consider the simplest matching 
procedure, namely the Landau prescription:
\begin{equation}
\left\langle [\cdots] \right\rangle_S = \left\langle [\cdots] \right\rangle_{QL_D}
\label{stoch-quant}
\end{equation}
subindex $S$ means \textit{stochastic}, subindex $Q$ stands for 
\textit{quantum mechanical} and subindex $L_D$ indicates spatial average
over a region with typical size $L_D$. 

We work under the hypothesis that the fluid is conformal and so the trace of the energy momentum tensor may be neglected. From the point of view of the formalism it is most convenient however to subtract it explicitly before computing the quantum expectation values. Neglecting the trace of the energy momentum tensor means we cannot include the effects of bulk viscosity in our analysis, we intend to address this issue in future works.
To reduce the complexity of the matching we assume that whenever a stochastic variable 
has a non-zero expectation value its fluctuations are negligible. This means that 
we shall treat the dimensionless temperature $\Upsilon$ and $\tilde u^0$ as true (but $L_D$-dependent) constants. 
In this way, from eqs. (\ref{II-d2-a}) and (\ref{u-lin}) we obtain our first nontrivial equation
\begin{equation}
3q_2 \Upsilon^4\left[1 + \frac{4}{3}\left\langle v_{(s)i}v_{(s)}^i\right\rangle_S\right] = \left\langle \hat T^{00}\right\rangle_{(QL_D)}
\label{T00-corr}
\end{equation}
and the symmetry requirement
\begin{equation}
\left\langle \tilde v^i \tilde v^j\right\rangle_S = \frac{1}{3}
\Delta^{ij} \left\langle v_{(s)i}v_{(s)}^i\right\rangle_S
\label{III-10}
\end{equation}
By the same logic, we shall not consider these quantities when looking at
second moments. The matching conditions we need then are
\begin{eqnarray}
\left\langle\left(\hat J^0\right)^2\right\rangle_S &=& \left\langle\left( \hat J^0\right)^2\right\rangle_{(QL_D)}
\label{III-11}\\
\left\langle \hat J^i \hat J^j\right\rangle_S &=&  \frac{1}{6}
\Delta^{ij} \left\langle\left\{\hat J^k,
\hat J_{k} \right\}\right\rangle_{(QL_D)} \label{III-12}\\
\left\langle \hat J^i \hat T^{0j}\right\rangle_S &=& 0 \label{III-13}\\
\left\langle \hat T^{0i} \hat T^{0j}\right\rangle_S &=& \frac{1}{6} 
\Delta^{ij} \left\langle\left\{ \hat T^{0k},\hat T_{0k}\right\}\right\rangle_{(QL_D)} \label{III-14}\\
\left\langle \hat {T}^{ij}_{(tls)}  \hat T^{kl}_{(tls)}\right\rangle_S &=& 
\frac{1}{2}\left\langle\left\{ \hat T^{ij}_{(tls)},
\hat T^{kl}_{(tls)}\right\}\right\rangle_{(QL_D)} \label{III-15}
\end{eqnarray}
where on both sides
\begin{equation}
\hat T^{ij}_{(tls)} = \hat T^{ij} - \frac{1}{3} \Delta^{ij}T^{k}_k
\label{III-16}
\end{equation}
For the last correlation, by demanding symmetry we can write
\begin{equation}
\frac{1}{2}\left\langle\left\{\hat T^{ij}_{(tls)},
\hat T^{kl}_{(tls)}\right\}\right\rangle_{(Q\lambda)}
= A\left(\Delta^{ik}\Delta^{jl} + \Delta^{il}
\Delta^{jk} - \frac{2}{3}\Delta^{ij}\delta^{kl}\right)
\label{III-17}
\end{equation}
with
\begin{equation}
A = \frac{1}{20} \left\langle\left\{\hat T^{ij}_{(tls)},
\hat T^{ij}_{(tls)}\right\}\right\rangle_{(Q\lambda)} 
\label{A-def}
\end{equation}
Observe that due to $\tilde u^0_{(s)} = 0$, the constraints on the hydrodynamic
variables are $\left\langle v_{(s)i} Z^i\right\rangle_S = \left\langle 
v_{(s)i} Z^{ij}\right\rangle_S = 0$, while $Z^i_i = 0$ identically, not just 
in expectation value. We shall not consider stochastic components in $Z^0$
nor in $Z^{0i}$. The previous matching relations give
\begin{equation}
\Upsilon^6\left\langle\left(eq_1 v_{(s)}^i + \Lambda e^2 \hat\tau Z^i\right)
\left(eq_1 v_{(s)}^j + \Lambda e^2 \hat \tau Z^j\right)\right\rangle_S =
\frac{1}{6}\Delta^{ij} \left\langle\left\{\hat J^k,
\hat J_{k} \right\}\right\rangle_{(QL_D)} \label{III-18}
\end{equation}
from whose trace we determine $\left\langle Z^k Z_k\right\rangle_S$ in terms of 
$\left\langle v_{(s)}^k v_{(s)k}\right\rangle_S$.

The current-EMT correlation vanishes because $\langle q_1\rangle_S = 
\langle q_1 v_{(s)}^i\rangle_S = \langle q_1 Z^i\rangle_S = 0$.
From the momentum density correlation we get
\begin{equation}
16 q_2^2\Upsilon^8\left[1+\left\langle v_{(s)}^k v_{(s)k}\right\rangle_S
\right]\left\langle v_{(s)}^l v_{(s)l}\right\rangle_S = \frac{1}{2}\left\langle\left\{ \hat T^{0k},
\hat T_{0k}\right\}\right\rangle_{(QL_D)} \label{T0i-corr}
\end{equation}
From eqs. (\ref{T00-corr}) and (\ref{T0i-corr}) we determine  
$\left\langle v_{(s)}^k v_{(s)k}\right\rangle_S$.
From the charge-charge correlation we get
\begin{equation}
\Upsilon^6 e^2\left[1 + \left\langle v_{(s)}^k v_{(s)k}\right\rangle_S\right]
\left\langle q_1^2\right\rangle_S = \left\langle \left(\hat J^{0}\right)^2
\right\rangle_{QL_D}\label{J0-corr}
\end{equation}
which determines the rms value of the charge imbalance. 

For the EMT-EMT correlation, we begin by observing that under the assumption of Gaussian fluctuations we have
\begin{eqnarray}
\left\langle v_{(s)}^i v_{(s)}^j v_{(s)}^k v_{(s)}^l\right\rangle_S 
&=& \frac{1}{9}\left\langle v_{(s)}^p v_{(s)p}\right\rangle_S^2
\left[\Delta^{ij}\Delta^{kl} + \Delta^{il}\Delta^{kj}
+ \Delta^{ik}\Delta^{jl}\right] \label{III-19}
\end{eqnarray}
and in particular
\begin{equation}
\left\langle v_{(s)}^i v_{(s)i} v_{(s)}^k v_{(s)}^l\right\rangle_S
= \frac{5}{9}\left\langle v_{(s)}^p v_{(s)p}\right\rangle_S^2 \Delta^{kl}
\label{III-20}
\end{equation}
Therefore the trace-free velocity correlation is
\begin{equation}
\left\langle \left(v_{(s)}^i v_{(s)}^j\right)_{tls}
\left(v_{(s)}^k v_{(s)}^l\right)_{tls}\right\rangle_S = 
\frac{1}{9}\left\langle v_{(s)}^p v_{(s)p}\right\rangle_S^2
\left[\Delta^{il}\Delta^{kj} + \Delta^{ik}\Delta^{jl}
- \frac{2}{3}\Delta^{ij}\Delta^{kl} \right]
\label{III-21}
\end{equation}
We therefore make the assumptions
\begin{eqnarray}
\left\langle v_{(s)}^i Z^{jk}\right\rangle_S &=& 0 \label{III-22}\\
\left\langle Z^{ij}Z^{kl}\right\rangle_S &=& 
\frac{1}{10} \left\langle Z^{pq}Z_{pq}\right\rangle_S
\left[\Delta^{ik}\Delta^{jl}
+ \Delta^{il}\Delta^{jk}
- \frac{2}{3}\Delta^{ij}\Delta^{kl} \right]
\label{III-23}
\end{eqnarray}
and get
\begin{equation}
A = \frac{16}{9} q_2^2 \Upsilon^8 
\left\langle v_{(s)}^p v_{(s)p}\right\rangle_S^2 
+ \frac{8}{125}q_2^2\hat\tau^2 \Upsilon^{10} \left\langle Z^{pq}Z_{pq}
\right\rangle_S \label{Tij-corr}
\end{equation}

Equations (\ref{T00-corr}), (\ref{III-18}), (\ref{T0i-corr}), (\ref{J0-corr}) 
and (\ref{Tij-corr}) determine the (stochastic) fluid quantities in terms 
of their quantum counterparts. Of them we shall calculate here only the ones that
determine the background, i.e. (\ref{T00-corr}) , (\ref{J0-corr}) and (\ref{Tij-corr}).

To end this section we must mention that it is straightforward to evaluate
the r.h.s. of correlation (\ref{III-14}) using the corresponding components of expr. 
(\ref{conf-KG-body}) together with the mode functions
(\ref{mode-asympt}). It is obtained that effectively
\begin{equation}
16 q_2^2\Upsilon^8\left\langle v^l_{(s)}v_{l(s)}\right\rangle \propto
\left\langle\left\{ \hat T^{0k},\hat T_{0k}\right\}\right\rangle_{(QL_D)} \ll 1
\label{T0i-corr-small}
\end{equation}
Therefore to study the instability we can consider only
\begin{equation}
A = \frac{8}{125}q_2^2\hat\tau^2 \Upsilon^{10}
\left\langle Z^{ij}Z_{ij}\right\rangle_{S} \equiv 
N^{ij}_{ij}(L_D)\label{nijij-xx}
\end{equation}
with $N^{ij}_{ij}$ given in expr. (\ref{Nijij-lambda}) and
from where we estimate the anisotropic pressure; and
\begin{equation}
\left\langle \hat T^{00}\right\rangle_{(QL_D)} = 3q_2\Upsilon^4
\label{T00-Temp}
\end{equation}
Eq. (\ref{J0-corr}), or the charge excess, becomes
\begin{equation}
\Upsilon^6 e^2 \left\langle q_1^2\right\rangle_S \equiv
\left\langle \left(\hat J^{0}\right)^2\right\rangle_{QL_D}
\label{J0-corr-final-b}
\end{equation}
with $\left\langle \left(\hat J^{0}\right)^2\right\rangle_{QL_D}$ given in expr. (\ref{J0J0-lambda-b}). The temperature
of the plasma will be calculated in Section \ref{weibel-prim}
using eqs. (\ref{T00-Temp}) and (\ref{J0-corr-final-b}) in the consistency relation 
(\ref{const-q1q2}).

\subsection{Anisotropic Background Pressure}\label{back-build-a}

In a certain coordinate system the anisotropic background pressure
tensor reads
\begin{equation}
Z^{ij} = \left(
\begin{matrix}
	Z_{0} + Z_{*} & 0 & 0\\
	0 & Z_{0} - Z_{*} & 0\\
	0 & 0 & -2Z_{0}
\end{matrix}
\right)
\end{equation}
We must find the statistics for $Z_{0}$ and $Z_{*}$, i.e., we must calculate
$\langle Z_{0}^2\rangle_S$, $\langle Z_{*}^2\rangle_S$ and 
$\langle Z_{0}Z_{*}\rangle_S$. We begin by asking the statistics to be 
invariant under the change of sign of $Z_{*}$, which is just a relabelling
of the coordinates, and then it must be that
$\langle Z_{0}Z_{*}\rangle_S=0$. We also demand that all representations have
the same statistics. If we define
\begin{eqnarray}
Z_{0} &=& -\frac{1}{2}\left(Z_{0}^{\prime} + Z_{*}^{\prime}\right)\label{zo-transf}\\
Z_{*} &=& -\frac{1}{2}\left(3Z_{0}^{\prime} - Z_{*}^{\prime}\right)\label{z*-transf}
\end{eqnarray}
then $Z^{ij}$ remains unchanged (modulo a rotation of coordinates). 
Demanding that the statistics is the same means $\langle Z_{0}^2\rangle_S =
\langle Z_{0}^{\prime 2}\rangle_S$ and $\langle Z_{*}^2\rangle_S =
\langle Z_{*}^{\prime 2}\rangle_S$. With the above correspondence between
representations we get that a necessary condition to have the same 
statistics is
\begin{equation}
\langle Z_{0}^2\rangle_{S} = \frac{1}{3}\langle Z_{*}^2\rangle_{S} \label{Zast-Z0}
\end{equation}
and then
\begin{equation}
\langle Z^{ij}Z_{ij}\rangle_S  = 12 \langle Z_{0}^2\rangle_{S}
\end{equation}
So together with eq. (\ref{nijij-xx}) we write
\begin{equation}
\langle Z_{0}^2\rangle_{S} = \frac{125}{96}\frac{1}{q_2^2\hat\tau^2
	\Upsilon^{10}} 
N^{ij}_{ij}(L_D) \, .\label{Z02-final}
\end{equation}
After $Z_0$ and $Z_*$ have been independently drawn according to this statistics, a final transformation
eqs. (\ref{zo-transf})-(\ref{z*-transf}) yields a new pair with  $Z_0>0$ and $Z_* < 3 Z_0$.

\section{Charged Scalar Field in a Flat FRW Universe}\label{scal-phi-sect}

As stated above, we assume the presence of a spectator charged scalar field coupled to gravity, which is in its vacuum state during Inflation. At the onset of the subsequent epoch to Inflation the vacuum becomes a particle 
state and in consequence a scalar plasma is formed. As the quantum state is rotationally invariant and the gravitational particle creation process is insensitive to the charge signs, the mean anisotropic pressure of 
the plasma and the charge excess will both be zero. However due to quantum fluctuations the rms deviations from those mean values are non-vanishing in any finite domain. As stated above the details of the calculation
to find the different expressions presented in this section can be found in Appendix \ref{ChScField}.

A charged scalar field is mathematically described by the complex
fields \cite{ItzZub80} $\mathbf{\Phi}(x^{\mu})$ and $\mathbf{\Phi}^{\dagger}(x^{\mu})$, which can be written in
terms of real fields as $\mathbf{\Phi}(x^{\mu}) = \left(\Phi_1 + i \Phi_2\right)/\sqrt{2}$
and $\mathbf{\Phi}^{\dagger}(x^{\mu}) = \left(\Phi_1 - i \Phi_2\right)/\sqrt{2}$,
where $\Phi_1$ and $\Phi_2$ commute and satisfy the Klein-Gordon equation for a massive real spin-0 
field arbitrarily coupled to gravity.
The Energy-Momentum tensor (EMT) in curved spacetime of each real component of the charged scalar field 
of dimensionless mass $\mu=m/H$  and coupled to gravity through a coupling constant $\xi$ is \cite{BirrDav82,ParkToms09}
\begin{eqnarray}
T_{\mu\nu}^{(\Phi)}  &=& \partial_{\mu}\Phi \partial_{\nu}\Phi 
-g_{\mu\nu}\left[\frac{1}{2} \left(1 - 4\xi\right)
g^{\alpha\beta}\partial_{\alpha}\Phi\partial_{\beta}\Phi 
+\frac{1}{2}\left(1-4 \xi\right) \mu^2\Phi^2 
-\frac{3}{2} \xi R\Phi^2\right]\nonumber\\
&+& \xi R_{\mu\nu}\Phi^2 - \xi\nabla_{\mu}\partial_{\nu}\Phi^2\, ;
\label{emt-KG-body}
\end{eqnarray}
The symbol $R_{\mu\nu}$ is the Ricci tensor and $R$ the scalar curvature. 
The values $\xi=0$ and $\xi =1/6$ 
correspond to the minimal and conformal coupling respectively. 
Rescaling the scalar field as
\begin{equation}
\Phi = \frac{\varPhi}{a} \label{conf-sf-body}\,,
\end{equation}
expr. (\ref{emt-KG-body}) becomes $T_{\mu\nu}^{(\Phi)}=a^{-2}\tilde T_{\mu\nu}^{(\Phi)}$ with
\begin{eqnarray}
\tilde T_{\mu\nu}^{(\Phi)} &=& \left(\partial_{\mu}\varPhi - \frac{\varPhi}{a}\partial_{\mu}a\right)
\left(\partial_{\nu}\varPhi - \frac{\varPhi}{a}\partial_{\nu}a\right) + \xi \tilde R_{\mu\nu}\varPhi^2 - \xi\nabla_{\mu}\partial_{\nu}\varPhi^2 \nonumber\\ 
&-&\left[\frac{1}{2} \left(1 - 4\xi\right)
\eta^{\alpha\beta}\left(\partial_{\alpha}\varPhi - \frac{\varPhi}{a}\partial_{\alpha}a\right)
\left(\partial_{\beta}\varPhi - \frac{\varPhi}{a}\partial_{\beta}a\right)\right. \nonumber\\
&&\left.+ \frac{1}{2}\left(1-4 \xi\right) \mu^2\varPhi^2 
-\frac{3}{2} \xi \tilde R\varPhi^2\right]\eta_{\mu\nu}
\label{conf-KG-body}
\end{eqnarray}
where $\eta_{\mu\nu}$ is the Minkowski metric tensor.

The electric four current is
\cite{ItzZub80}
\begin{equation}
J_{\mu}\left(x^{\nu}\right) = - e \left[\Phi_1 \left(x^{\nu}\right)
\partial_{\mu}\Phi_2 \left(x^{\nu}\right) - \Phi_2\left(x^{\nu}\right)
\partial_{\mu}\Phi_1\left(x^{\nu}\right) \right]\label{ell-curr-2-body}
\end{equation}
After replacing (\ref{conf-sf-body}) for each field we have
that $J_{\mu}=a^{-2}\tilde J_{\mu}$ with
\begin{equation}
\tilde J_{\mu} = - e \left[\varPhi_1  \left(\partial_{\mu}\varPhi_2 - \frac{\varPhi_2}{a}\partial_{\mu}a\right)
- \varPhi_2  \left(\partial_{\mu}\varPhi_1 - \frac{\varPhi_1}{a}\partial_{\mu}a\right)\right]
\label{ell-curr-3-body}\,.
\end{equation}

\subsection{Computing the Energy Density and Noise Kernels}

To compute the energy density and the noise kernels of the quantum scalar gas needed to build the 
fluid description, we consider that during Inflation the scalar field is in its vacuum state and after 
the transition to Reheating only adiabatically regularized modes will be present \cite{ParkToms09}.
This means that observers in the local vacuum state will detect a bath of particles built up from 
super-horizon modes.

In Section \ref{back-build} we construct the correspondence between a quantum gas and its fluid description assuming the simplest possible matching procedure. We only need three functions in order to set the background of the hydrodynamic theory, namely the VEV $\langle \hat T^{00}\rangle_Q$  and the traces 
$\langle\{\hat T^{ij}(\bar r_1), \hat T^{ij}(\bar r_2)\}\rangle_Q$ and
$\langle\{\hat J^{0}(\bar r_1),\hat J^{0}(\bar r_2)\}\rangle_Q$ related to the anisotropic 
pressure and the electric charge excess respectively. In the next subsection we compute these functions.

\subsubsection{Computing $\langle \hat T^{00}\rangle$} \label{sbs-T00-vev}

\begin{figure}[t]
\centering
\includegraphics[width=10cm]{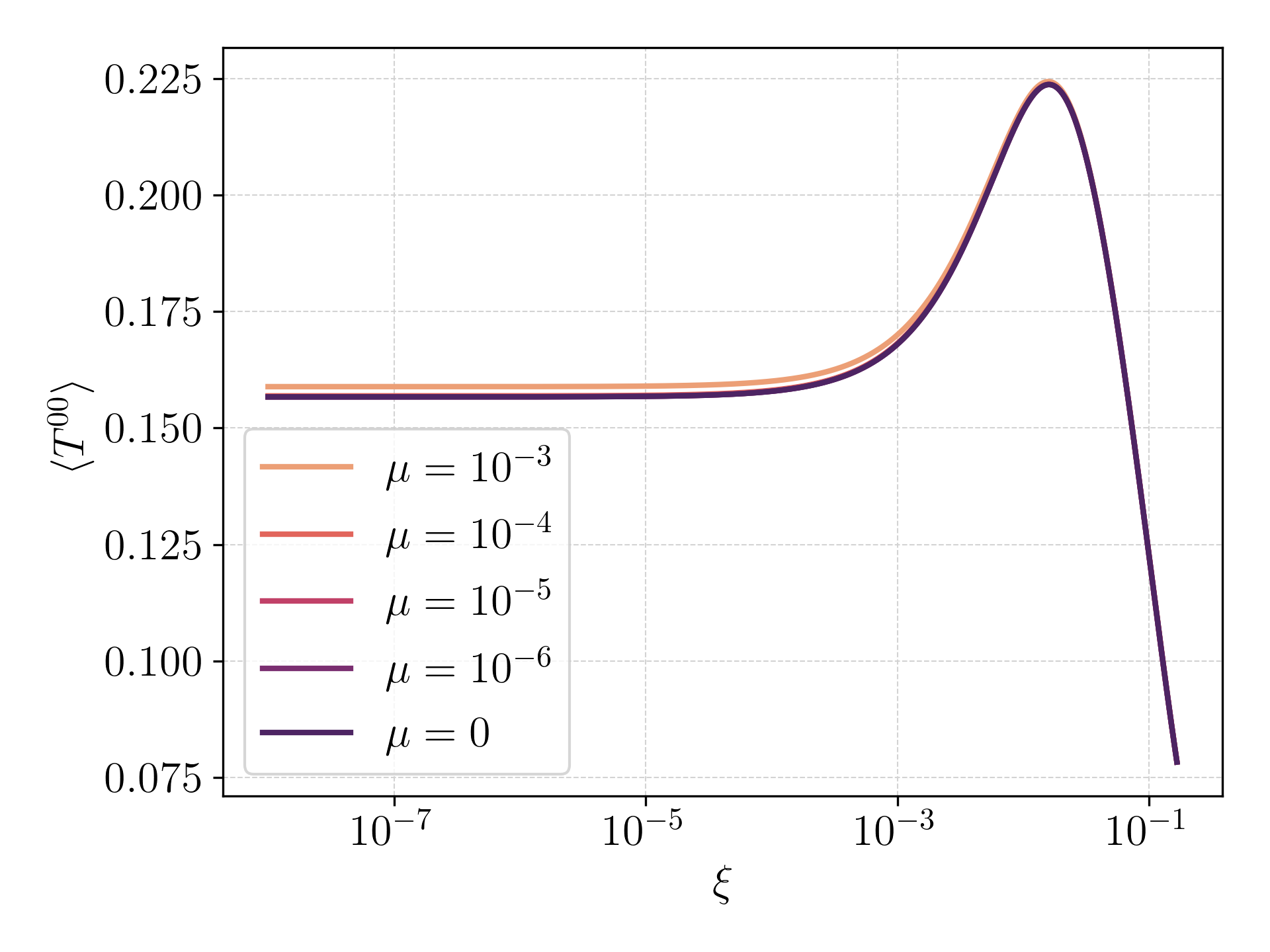} \\
\caption{\label{toofull} We show the VEV of the 00-component of the energy-momentum tensor vs. the coupling to gravity. This quantity defines the energy density of the fluid. The integration of eq. (\ref{T00-exact}) was performed up to $k=1$ and for different values of the dimensionless masses $\mu$ shown in the legends. We observe that the result
	is quite independent of this parameter.}
\end{figure}

In this section we evaluate the VEV of $\hat T_{00}$
at $\eta=0$ from expr. (\ref{conf-KG-body}). Recall that the factor $2$ is due to the fact that a 
complex scalar field consists of two real fields, and expr. (\ref{conf-KG-body}) corresponds
to only one of those fields. Replacing the expansion (\ref{phi-fourier}) and evaluating the 
expectation value in an adiabatic vacuum state we obtain
\begin{eqnarray}
\left\langle\tilde T_{00}\right\rangle
&=&  \frac{2^{3/2}}{\pi^{1/2}} \int_0^1 dk~ k^2
\left\{\frac{1}{2}\left[\left\vert\varphi^{\prime}_k
-\varphi_k\right\vert^2 + \left( k^2  +  \mu^2 \right)
\left\vert\varphi_k\right\vert^2 \right]
+ \xi  \left[1- 2\left( k^2  + \mu^2 \right)
\right]\left\vert\varphi_k\right\vert^2
\right.\nonumber\\
&&\left. - 2\xi \mathrm{Re}\left[\varphi_k^{\ast}
\left(\varphi_k''- 3\varphi'_k \right)\right]
- 24 \xi^2  \left\vert\varphi_k\right\vert^2\right\}
\label{T00-exact}
\end{eqnarray}
We numerically integrate this expression and in Figure \ref{toofull} we show the value of $\langle \tilde T^{00} \rangle$ as a function of $\xi$ for several values of the field mass. 
In the case of small masses $\mu\ll1$ this VEV is insensitive to $\mu$, but it shows a strong dependence on 
the coupling to gravity $\xi$. 

\subsubsection{Anisotropic pressure noise kernel $N_{ij}^{ij}( x, y)$}
\label{Nijij-subsec}

In this section we compute the traceless part of the pure spatial sector of the energy-momentum tensor which carries on the information of the anisotropic pressure. The corresponding noise kernel is the 
vacuum expectation value of the contracted symmetric two-point function
\begin{eqnarray}
N^{ij}_{ij}(x,y) &=& \left\langle\left\{ \tilde T^{ij}_{(tls)}(x),
\tilde T_{(tls)ij}(y)\right\}\right\rangle \label{nijij-1}
\end{eqnarray}
with $\tilde T_{(tls)}^{ij}(x) = \hat T^{ij} - (1/3) \Delta^{ij}\hat T^{k}_k$. Replacing
the Fourier mode decomposition of the scalar field we obtain

\begin{eqnarray}
N^{ij}_{ij}(\bar x,\bar y) &=& 8\int\frac{d^3p}{(2\pi)^{3/2}}\frac{d^3q}{(2\pi)^{3/2}} 
\cos\left[(\bar p+\bar q).(\bar x-\bar y)\right]
\left\vert\varphi_p\right\vert^2 \left\vert\varphi_q\right\vert^2
\left\{\left(1-2\xi\right)^2 \left[p^2q^2 + \frac{1}{3}
(\bar p.\bar q)^2\right]
\right.\nonumber\\
&& \left. - \frac{8}{3} \xi\left(1-2\xi\right) 
(\bar p.\bar q)(p^2 + q^2)
+ 4\xi^2 \left[(\bar p.\bar q)^2 - \frac{1}{3}p^2q^2 
+ \frac{1}{3}(p^4 + q^4) \right] \right\}
\label{nijij-10}
\end{eqnarray}
To evaluate the momentum integrals we can retain only the leading term in the expansion of the scalar field
mode functions (see Appendix \ref{ChScField}) and write
\begin{equation}
\varphi_k \simeq -\frac{\pi^{1/2}}{2} 
\frac{i e^{ik}}{\sin(\nu\pi)}
\frac{1}{\Gamma(1-\nu)} \frac{2^{\nu}}{k^{\nu}}
\label{mode-asympt-body}\,.
\end{equation}
with $\nu = (3/2)\sqrt{1 - (16/3)\xi}$.
Morover we consider two asymptotic regimes, $r \lesssim 1$ and $r \gtrsim 1$, and interpolate between both
regions. For $r \gtrsim 1$ we obtain
\begin{equation}
N^{ij}_{ij}(r\geq 1) \simeq 
\frac{16}{3\pi} \frac{1}{r^{10-4\nu}}
\label{nijij-pq-only-c2}
\end{equation}
while for $r \lesssim 1$ we have
\begin{eqnarray}
N^{ij}_{ij}(r\lesssim 1) &\simeq&  \frac{4}{2\pi}
\left\{\frac{14}{9} -\frac{91}{450} r^2 \right\}
\label{nijij-pq-only-f2b}
\end{eqnarray}

We must now apply a spatial average over a scale of interest $L_D$ to filter the kernel $N_{ijij}(r)$. 
We consider a top-hat window function and write
\begin{equation}
N^{ij}_{ij}(L_D) = \frac{3}{4\pi \lambda^3}
\int_0^{\lambda} d^3r N_{ijij}(r)
= \frac{3}{\lambda^3}\left[
\int_0^{r_c} dr~ r^2~ N_{ijij}(r)
+ \int_{r_c}^{\lambda} dr~ r^2~ N_{ijij}(r) \right]
\label{Nijij-filtered}
\end{equation}
with $r_c \sim 1$ the coordinate at which both parts of
$N_{ijij}(r)$  cross.
Replacing exprs.  (\ref{nijij-pq-only-c2}) and (\ref{nijij-pq-only-f2b}) into (\ref{Nijij-filtered})
we have that the filtered kernel reads
\begin{eqnarray}
N^{ij}_{ij}(L_D)  &\simeq & \frac{12}{L_D^3}\left[\frac{a}{3}r_c^3 - \frac{b}{5}r_c^5  
+  \frac{d}{4\nu-7}\left(\frac{1}{r_c^{7-4\nu}}
- \frac{1}{L_D^{7-4\nu}} \right) \right] \label{Nijij-filtered-3}\,.
\end{eqnarray}
Taking $r_c\sim 1$, we finally have for scales larger than the horizon at the end of 
Inflation
\begin{eqnarray}
N^{ij}_{ij}(L_D)  &\sim & \frac{28}{3\pi}\frac{1}{L_D^3}
\label{Nijij-lambda}\,.
\end{eqnarray}

\subsubsection{Charge excess noise kernel 
$\mathcal{J}_{00}(\bar x,\bar y)$}\label{J00-subsec}

To compute the rms value of the electric charge we define a kernel 
for the electric current in a similar way as we did in the previous section, 
namely the VEV of the symmetric two-point function
\begin{equation}
\mathcal{J}_{00}(x,y) = \left\langle 0 \right\vert
\left\{\tilde J_{0}(x),\tilde J_{0}(y)\right\}\left\vert 0 \right\rangle
\label{J00-NK}
\end{equation}

Replacing expr. (\ref{phi-fourier}) and (\ref{ell-curr-3-body}) into (\ref{J00-NK}) we obtain that 
the only terms which contribute to the kernel are
\begin{eqnarray}
\mathcal{J}_{00}(x,y) &=& 2 e^2 \int\frac{d^3p}{(2\pi)^{3/2}}
\int\frac{d^3q}{(2\pi)^{3/2}} \cos\left[(p+q)(x-y)\right] 
\left\vert\varphi_{p}\varphi'_{q} - \varphi_{q}\varphi'_{p}\right\vert^2 \label{j001}
\end{eqnarray}
To estimate the momentum integrals we 
again consider two regimes $r \lesssim 1$ and $r \gtrsim 1$.
For $r\lesssim 1$ we obtain
\begin{eqnarray}
\mathcal{J}_{00}(r\lesssim 1) &\simeq& \frac{e^2}{12\pi} \frac{1}{\xi}
\left[1-\frac{r^2 }{12}\right]\,, \label{J00smlr-b}
\end{eqnarray}
while for $r\gtrsim 1$ we get
\begin{eqnarray}
\mathcal{J}_{00}(r\gg 1) &\simeq& -\frac{e^2}{2\pi} \frac{1}{\xi}  \frac{1}{r^{10-4\nu}}
\label{J00lrgr-b}\,.
\end{eqnarray}

We now proceed with the average over a scale $\lambda$ using again a top-hat window function. 
Since $r_c$ is the coordinate where both expressions match we have
\begin{equation}
\mathcal{J}_{00}(L_D) = \frac{3}{4\pi\lambda^3}
\int_0^{L_D} d^3r\,  \mathcal{J}_{00}(r) =
\frac{3}{L_D^3}\left[\int_0^{r_c}dr~r^2\mathcal{J}_{00}(r)
+ \int_{r_c}^{L_D} dr~r^2\mathcal{J}_{00}(r)\right]\label{J0J0-filter}\,.
\end{equation}
Finally assuming $r_c \sim 1$, for large $L_D$ we obtain
\begin{eqnarray}
\mathcal{J}_{00}(L_D)  &\simeq& 8\frac{e^2}{\pi \xi}
\frac{1}{L_D^3}
\label{J0J0-lambda-b}\,.
\end{eqnarray}

\section{Primordial Weibel Instability}\label{weibel-prim}

In this section we study the dispersion relation from eq. (\ref{disp-rel-pm}). We begin by building the functions that describe the background fluid, $Z_{0zz}'$, $q_{10}$ and $q_{20}$, and computing the 
parameters $\hat\tau$ and $\Upsilon_0$. From eqs. (\ref{Z02-final}) and (\ref{Nijij-lambda})  we read
\begin{equation}
Z_0 = \sqrt{\langle Z_{0}^2\rangle_{S}} \simeq \frac{1.96}{q_{20}\hat\tau
	\Upsilon_0^{5} L_D^{3/2}}  \, .\label{Z0-final-body}
\end{equation}
and from the relation (\ref{Zast-Z0}) we have
\begin{equation}
Z_* \simeq \frac{3.40}{q_{20}\hat\tau
	\Upsilon_0^{5} L_D^{3/2}} \label{Zast-final-body}
\end{equation}
which is consistent with the constraint $Z_{*} < 3 Z_0$ we assume before.

\begin{figure}[ht]
\centering
\includegraphics[width=10cm]
{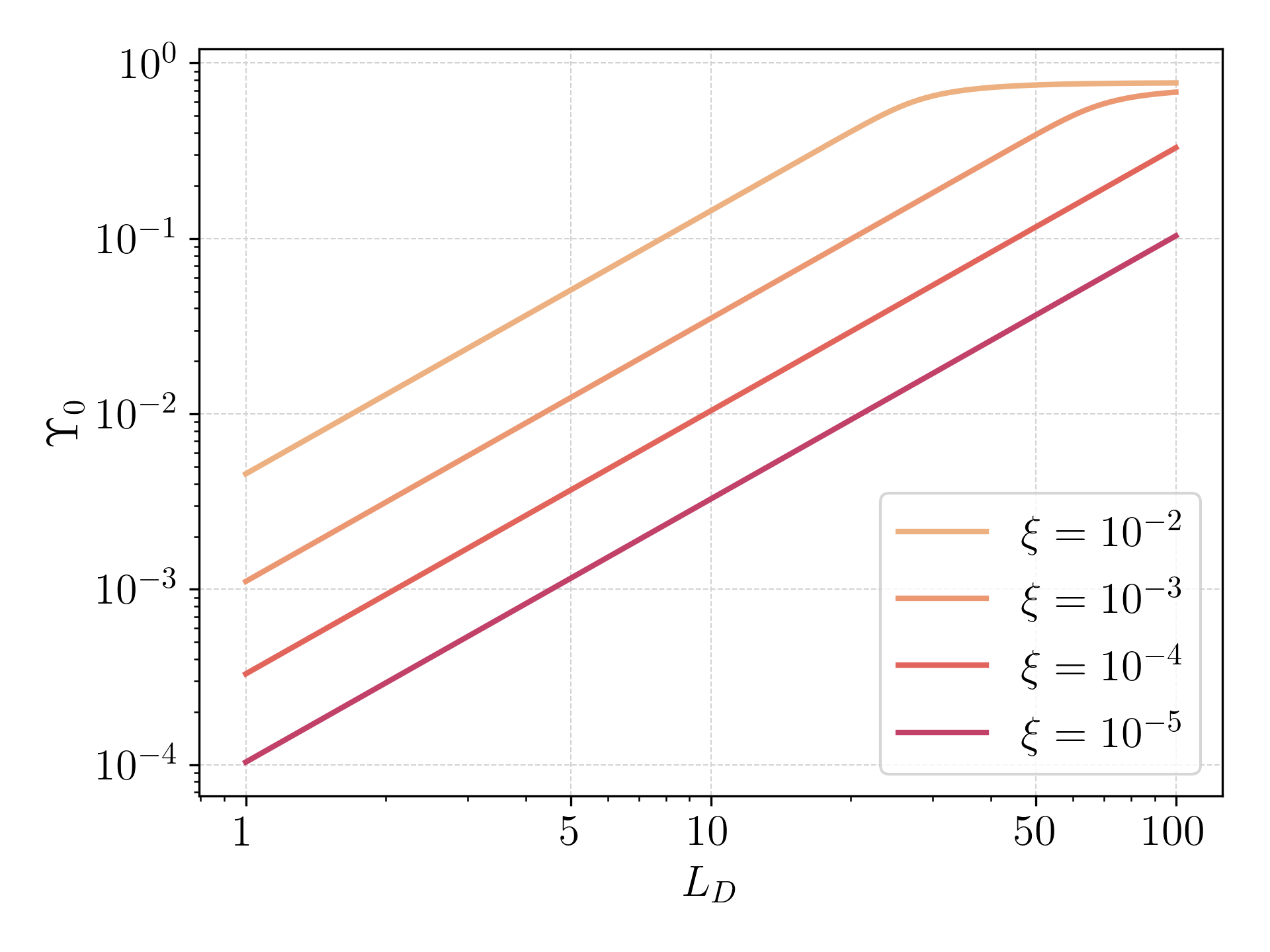}
\caption{Dimensionless temperature $\Upsilon_0$ given from the only real root of eq. (\ref{temp-from-root}) for several values of the coupling to gravity $\xi$. We see that in general it is higher than the dimensionless mass $\mu$ of the possible scalar field candidates (see the main text and Figure \ref{toofull} also).}
\label{tempvsL}
\end{figure}

To estimate $q_{10}^2$ we use eq. (\ref{J0-corr-final-b}) together with eq. (\ref{J0J0-lambda-b}) to
obtain
\begin{equation}
q_{10}^2 = \left\langle q_{10}^2\right\rangle_s \simeq \frac{8}{\pi \xi \Upsilon_0^6 L_D^3}
\label{q10-quant-stoch}
\end{equation}
The number density $q_{20}$ can be obtained from eq. (\ref{T00-Temp}) and reads
\begin{equation}
q_{20}  = \frac{\left\langle T^{00}\right\rangle}{3 \Upsilon_0^4}\, . \label{q20-tmunu}
\end{equation}
The numerical value of the numerator in the r.h.s. of this equality must be read from Figure \ref{toofull}
according to the chosen value of $\xi$.
The temperature in eqs. (\ref{q10-quant-stoch}) and (\ref{q20-tmunu}) can be 
fixed from the constraint (\ref{const-q1q2}) which becomes a polynomial equation
for $\Upsilon_0$, namely
\begin{equation}
\Upsilon_0^8 +\frac{2\pi^3}{\xi L_D^3}\Upsilon_0^2-\frac{\pi^4\langle T^{00}\rangle^2}{36} = 0\,. \label{temp-from-root}
\end{equation}
The dimensionless temperature $\Upsilon_0$ turns out to be a function of $\xi$ and $L_D$. This equation has only one real positive root shown in Figure \ref{tempvsL} as a function of the averaging domain size $L_D$ for different values of $\xi$.

A simple comparison between the values of the temperatures from Figure \ref{tempvsL} and the values of the masses in Figure \ref{toofull} shows that $\Upsilon_0 \gg \mu$ is satisfied 
for a wide range of small masses that include Axion-Like-Particles (ALPs) 
\cite{Marsh16,BauGreWa16,HOTW17,GhSa20}. In addition, since the scalar curvature
of a Universe that expands as radiation-dominated one is vanishing, the fluid of scalar particles 
can indeed be considered as conformal \cite{CalHu08}.

\begin{figure}[t!]
\centering
\includegraphics[width=10cm]
{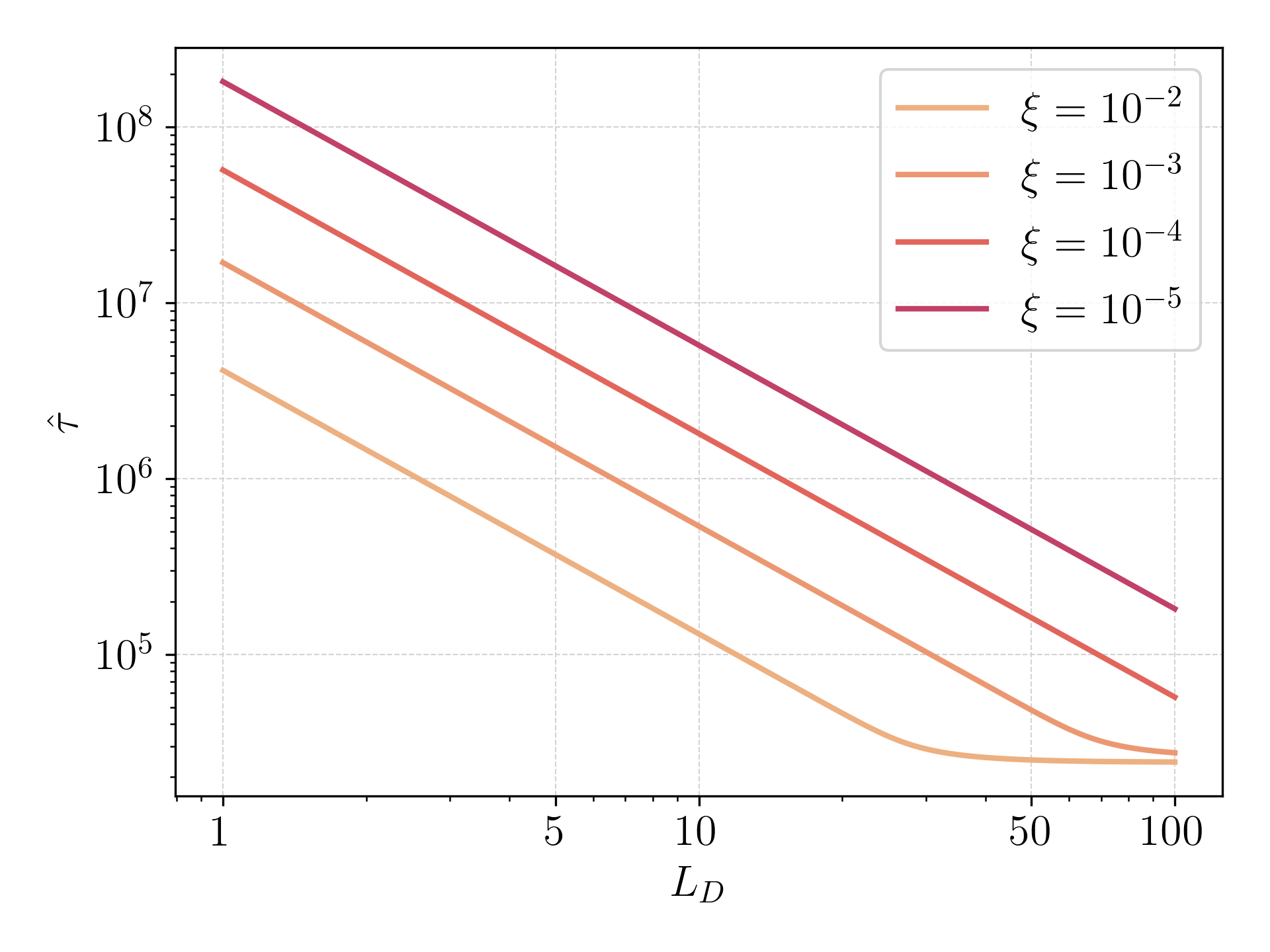}
\caption{Relaxation time $\hat\tau$ given by the eq. (\ref{tau-choice}) as a function of the averaging scale $L_D$ for several values of the gravity coupling $\xi$. We read the temperature in (\ref{tau-choice}) from eq. (\ref{temp-from-root}) and Figure \ref{tempvsL}. Note that for any choice of the free parameters $\hat\tau>10^4$.}
\label{tau-vs-L}
\end{figure}

The remaining parameter, the relaxation time $\hat\tau$, is determined by the scattering cross section of 
the self-interaction between the fluid particles. Assuming that the short range electromagnetic 
interactions provide the main scattering mechanism, we get
\begin{equation}
\hat\tau \simeq \frac{1}{\alpha_{fs}^2\Upsilon_0} \label{tau-choice}\,,
\end{equation}
with $\alpha_{fs} = e^2/4\pi \simeq 1/137$. This relaxation time sets the lifespan of the anisotropic pressure background, i.e. the time during which the instability can evolve. In Figure \ref{tau-vs-L} we show the dimensionless relaxation time $\hat \tau$ as a function of the averaging domain size $L_D$.

Our one-fluid model of the instability is valid during the intermediate stage between the end of Inflation and the establishment of the bulk radiation content of the Universe that sets the end of the so-called Reheating era and the beginning of the radiation-dominated era \cite{Loz19,LozAmin16}. Afterwards the charges will strongly couple to the bulk radiation and the relaxation time (\ref{tau-choice}) will no longer be the main mechanism of relaxation or thermalization due to the inclusion of the interactions with radiation. Then the optimal scenario for the development of the instability is the one for which the relaxation time is of the order of or larger than the duration of Reheating.

\subsection{Study of the Unstable Modes}

Now that we have all the parameters of the model, we study the unstable solutions of the dispersion 
relation (\ref{disp-rel-pm}). In principle we have two sets of solutions, one corresponding to
the $s_+$ modes and the other to the $s_-$ modes. We numerically solve the eq. (\ref{disp-rel-pm}) using the set of parameters defined at the beginning of this section. 

\begin{figure}[b!]
\centering
\includegraphics[width=10cm]
{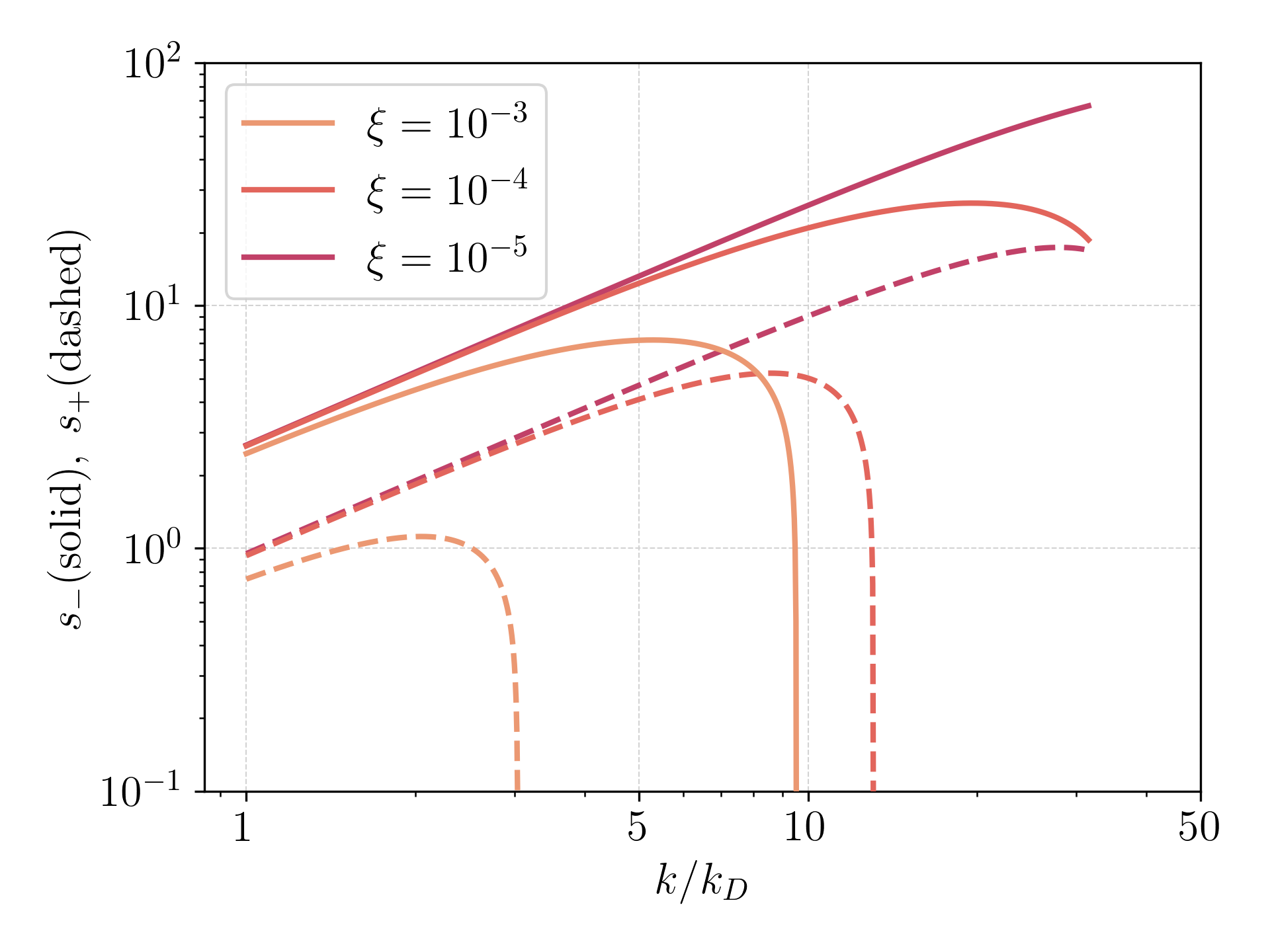}
\caption{Instability exponents $s_+$ and $s_-$ as functions of the scale $k/k_D$, with $k=\kappa/\hat\tau$, $k_D=2\pi/L_D$ and $L_D\simeq 20$, as a representative averaging domain size, for different values of the gravity coupling $\xi$. We choose $\theta = \varphi=0$ for the propagation direction. The relation $s_- > s_+$ is satisfied for any choice of the parameters.} 
\label{sMassMenos}
\end{figure}

We start analyzing the strength of the instability regarding both the magnetic field component (related to the $\pm$ modes) and the orientation of the propagation direction with respect to the anisotropic pressure background (related to the angles $\theta$ and $\phi$). We find that $s_-(k) > s_+(k)$ for any choice of $\xi$ and $L_D$ regardless of the scale $k$ and its orientation $(\theta,\phi)$. In Figure \ref{sMassMenos} we show $s_+$ and $s_-$ as a function of $k/k_D$, with $k=\kappa/\hat\tau$, $k_D=2\pi/L_D$ and $L_D\simeq 20$, for three different values of gravity coupling as representative examples.

\begin{figure}[b!]
\centering
\includegraphics[width=10cm]
{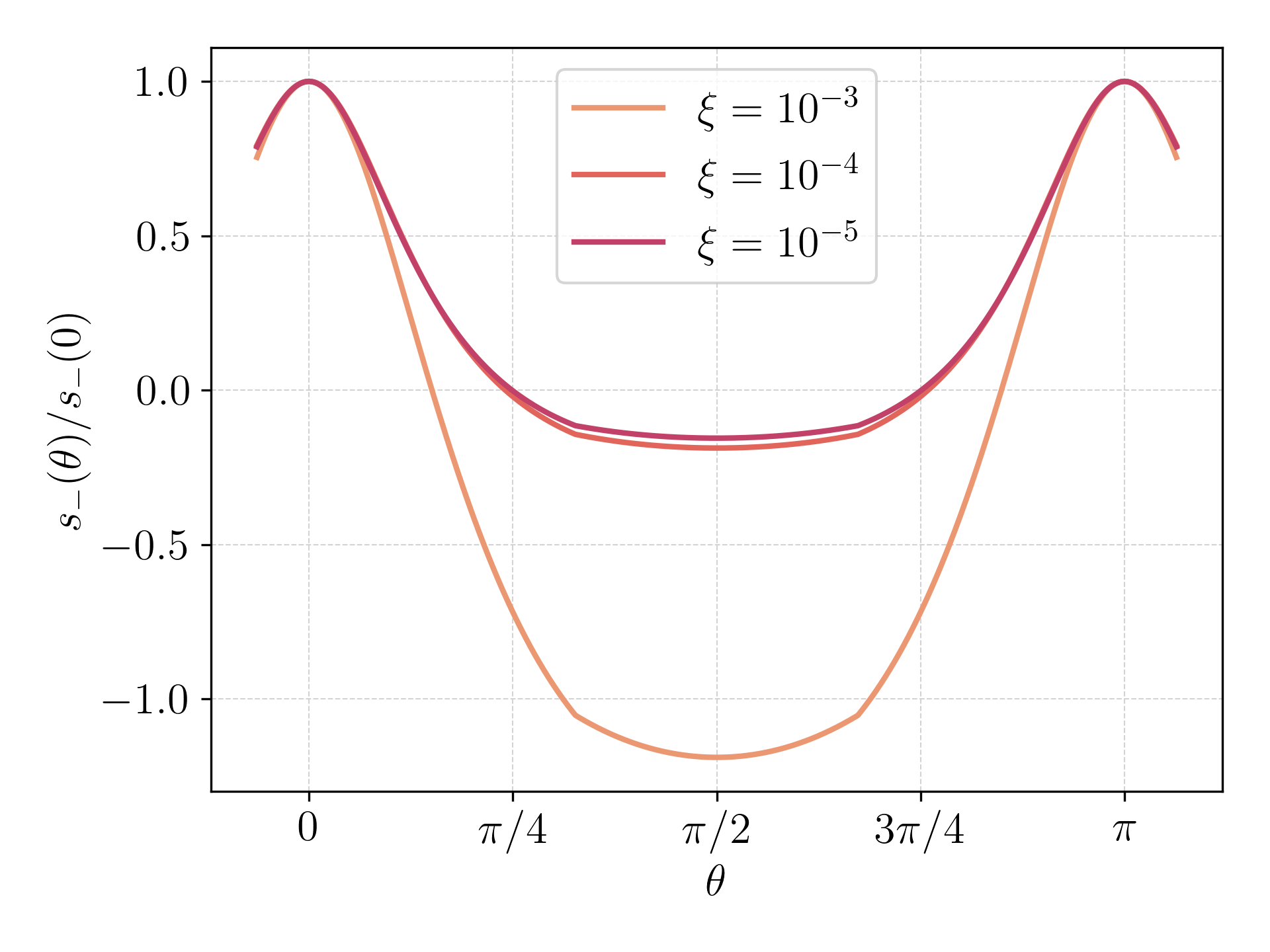}
\caption{Normalized instability exponent $s_-$ as a function of $\theta$ regarding the orientation between the propagation direction and the anistropic pressure tensor, for different gravity coupling values. As a representative example we take $L_D=60$, $k=2\pi/L_D$ and $\varphi = 0$. We observe that the largest values of the instability exponent $s_-$ arises for $\theta=0$ or $\theta=\pi$, i.e. when the propagation direction is aligned to the orientation of the anisotropic pressure tensor. We also see a range of angles for which the perturbations are stable.} 
\label{svstheta}
\end{figure}

In view of these results we focus on the $s_-$ solution. The dependence of the instability exponent $s_-$ with respect to the orientation of the propagation direction is shown in Figure \ref{svstheta}. In particular we show the values of $s_-$ as a function of $\theta$ with fixed $\varphi=0$ for three different gravity coupling values and, in this example we also fix the scale $k=k_D$ and $L_D\simeq 60$. Clearly the alignment (anti-alignment) of the propagation direction and the orientation of the anisotropic pressure tensor, i.e. $\theta=0$ ($\theta=\pi$), produces the strongest instabilities. Hereafter we focus on the $s_-$ modes with $\theta=\varphi=0$.

\subsubsection{Conformal and unstable evolution of the magnetic field}

To compute the evolution of the amplitude of the magnetic field we begin by rewriting the 
scale factor of the universe $a(\eta)$ in terms of \textit{e-folds} according
to the conventions and definitions in \cite{LozAmin16}. We assume that Reheating begins at $\eta=0$ and the scale factor, normalized to $a(\eta=0)=1$, evolves as in a radiation-dominated universe since the beginning of Reheating \cite{Loz19,LozAmin16}. In consequence the scale factor reads
\begin{equation}
a\left(\eta\right)= \left(1+\eta\right) =e^{N_{\eta}}, \label{efolds-def}
\end{equation}
with $N_\eta$ the number of \textit{e-folds} at the time $\eta$. The evolution of the magnetic field for $\eta < \hat\tau$ will be given by
\bea
\frac{B(\eta)}{B_I}\sim e^{\sigma_-(\bm k)\;\eta} e^{-2N_{\eta}}\,,\label{bbi}
\tea
where $B_I$ is the magnetic fields at the end of Inflation.

We consider Reheating scenarios whose duration $\eta_{f}$ satisfies $\eta_{f} \leq \hat \tau$. In these cases the instability is operative during all the Reheating stage. From Figure \ref{tau-vs-L} we may estimate that $\eta_{f} \leq 5\times 10^4$ which implies that the number of \textit{e-folds} at the end of Reheating, i.e. $N_{\eta_f}=N$, must fulfill $N \lesssim 10$.

Our aim is to compute the evolution of the magnetic field until the end of Reheating from the expression (\ref{bbi}). In order to determine all the hydrodynamics parameters we first choose the averaging domain size $L_D$ such that the average is performed over all the scales that are inside the horizon at $\eta_f$. Indeed for a given $N$ we have $L_D=e^N$ which coincides with the comoving horizon at $\eta_f$ implying that $k_f=2\pi a_fH_f=k_D$.

\begin{figure}[th!]
\centering
\includegraphics[scale=0.95]
{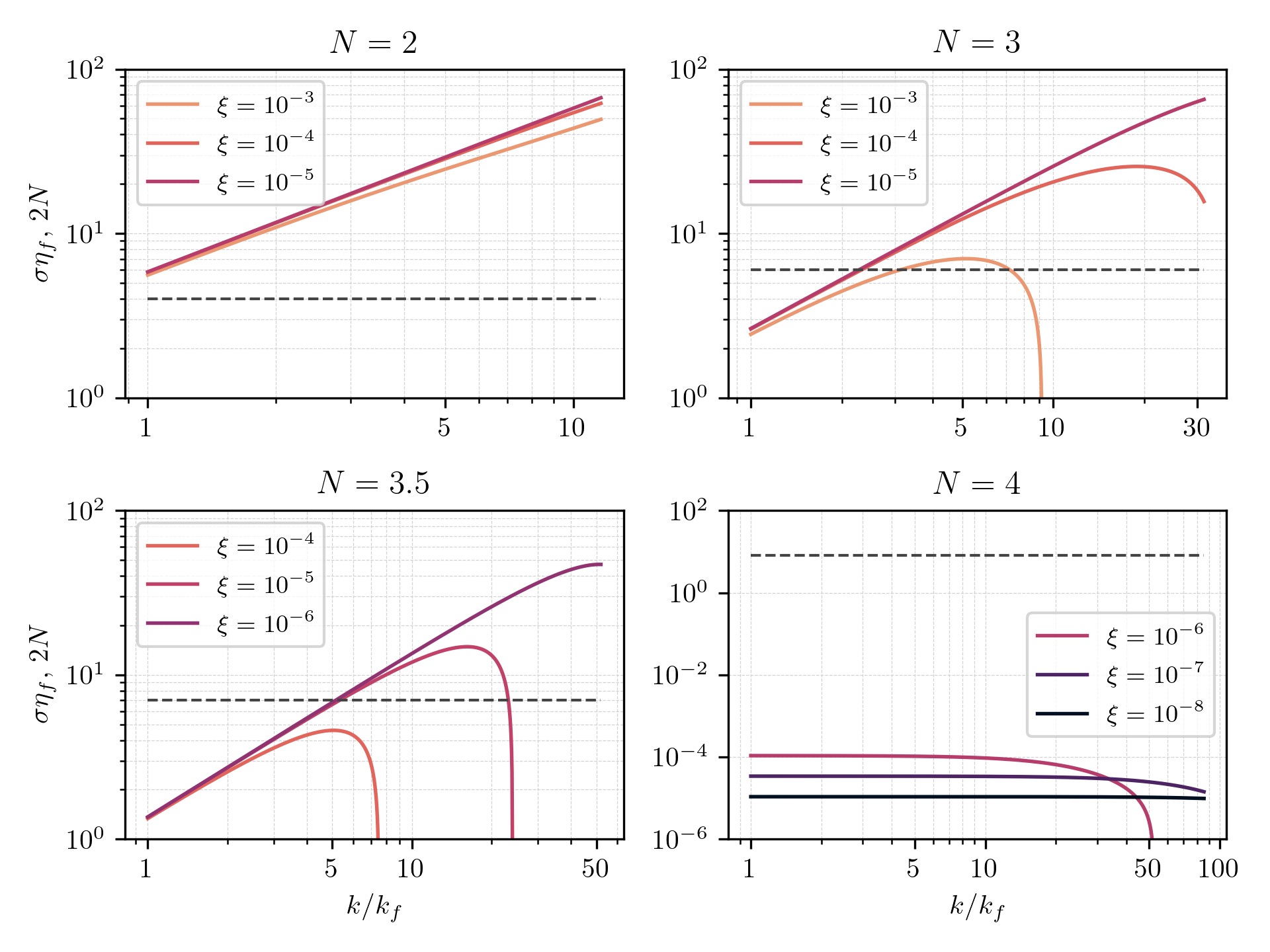}
\caption{Instability exponent spectrum at the end of Reheating. Solid lines corresponds to the instability exponents $\sigma_-\eta_f=s_-\eta_f/\hat\tau$ for different gravity coupling values and the dashed line is the conformal dilution exponent $2N$. The scale $k$ ranges from the comoving horizon at the end of Reheating $k=k_f$ to the comoving Inflation horizon $k=1$. The average of fluctuations, that determines the magneto-hydrodynamic parameters involved in $\sigma_-$, is performed over a region of the size of the horizon at the end of Reheating, i.e. $k_D=2\pi/L_D=k_f$. We observe that for $N\lesssim4$ the unstable exponential growth compensates the conformal dilution or even produces a net amplification depending on the gravity coupling $\xi$. As $N$ gets larger the instability can only overcomes the conformal dilution for lower values of $\xi$. For $N\gtrsim 4$ however the conformal dilution dominates for all 
	possible $\xi$.} 
\label{results_fig}
\end{figure}

In Figure \ref{results_fig} we compare the instability exponent spectrum ($\sigma_-(k)\eta_f$) and the conformal dilution exponent ($2N$) in (\ref{bbi}) as functions of the wavenumber scale $k/k_f$ with $k_f<k<1$ for different values of $\xi$ and several Reheating scenarios determined by $N$. We observe that for $N < 4 $, the magnetic field instability overcomes the conformal dilution for different ranges of scales $k$ depending on the value of $\xi$. For shorter Reheating periods with larger values of $\xi$ the instabilities can surpass the dilution, while for longer Reheating stages smaller values of $\xi$ are needed to achieve a net amplification. Nonetheless there is limit $N\sim 4$ for which the conformal dilution prevails for any value of $\xi$ and $k$.

\subsection{Bonus Track: Magnetic Helicity}\label{magnetic-helicty}

The features of the evolution of magnetic fields are not only determined by the Maxwell and 
fluid equations, but also by the topological properties of the field. They are encoded in the 
\textit{magnetic helicity} (MH), which is defined as the volume integral of the 
scalar product between the vector potential and the magnetic field \cite{BF84},
\cite{biskamp93}, \cite{biskamp03}. It is interpreted as the signed number of twists due 
to individual field lines as well as due to the interlinkages of different lines. 

Magnetic helicity is produced by the magnetic field creation mechanism itself, and  
it is not simple to envisage a primordial magnetogenesis process that creates fields with 
non-trivial topology.  A possibility is the electroweak epoch, when violation of parity and 
charge+parity occurred \cite{corn-97,vach-91,vach20,JackPi-00,ZhaFeVach17}, and also the 
inflationary epoch \cite{CapSor14,CarGuzzSor18} due to couplings of the electromagnetic field
with other fields as the axion. 
In addition stochastic magnetogenesis mechanisms can also induce fields with non-vanishing rms values of MH
\cite{calkan14,kmlk-14}.

In this Sub-section we build the expressions to compute the amplification that an rms magnetic 
helicity would undergo due to the instability studied in this Section. 
Here we drop the explicit prime symbol in order to keep a minimal notation. We saw that
we can build a basis with the three mutually orthonormal vectors 
$\{\bar\varepsilon_+(\mathbf{k}),\bar\varepsilon_-(\mathbf{k}),\hat{\mathbf{k}}\}$, where the 
first two are the eigenvectors associated to the eigenvalues $\lambda_+$ and $\lambda_-$ respectively and the third is the propagation direction. We assume that $\{\bar\varepsilon_+(\mathbf{k}),\bar\varepsilon_-(\mathbf{k}),\hat {\mathbf{k}}\}$ is right-handed, which means that the vectors satisfy 
$\bar\varepsilon_+(\mathbf{k})\times \bar\varepsilon_-(\mathbf{k})=\hat {\mathbf{k}}$.
The eigenvectors also satisfy $\bar\varepsilon_{\pm}(-\hat {\mathbf{k}}) 
= \bar\varepsilon_{\mp}(\hat {\mathbf{k}})$ under parity transformation of the propagation direction. 
To preserve the correspondence between eigenvectors and eigenvalues, the 
latter must transform as $\lambda_{\pm}(-\hat {\mathbf{k}}) = \lambda_{\mp}(\hat{\mathbf{k}})$ under parity.

Some time after the beginning of the evolution, when the decaying solutions have 
substantially faded away, each field of the vector sector can be written 
as a linear superposition of the unstable solutions according to
\begin{equation}
f^i\left(\mathbf{r},\eta\right) = \int \frac{d^3 \mathbf{k}}{(2\pi)^{3/2}} 
e^{i \mathbf{k} .\mathbf{r}}
\left[ e^{s_+(\mathbf{k}) \eta} f_+(\mathbf{k}) \varepsilon_+^i(\mathbf{k})
+ e^{s_-(\mathbf{k}) \eta} f_-(\mathbf{k}) \varepsilon_-^i(\mathbf{k})\right]\,.
\end{equation}
with $f_+(-\mathbf{k})=f_-^*(\mathbf{k})$ such that $f^i\left(\mathbf{r},\eta\right)$ is real. 
Since we analyze the dynamics only after some time of the evolution, 
we cannot consider an initial value problem. Instead we assume that
\begin{equation}
\left\langle f_+(\mathbf{k})\right\rangle = 0
= \left\langle f_-(\mathbf{k})\right\rangle\,,
\end{equation}
where the angular brackets denote ensemble average, and that for not-so-long times,
\begin{equation}
e^{s_+(\mathbf{k}) \eta} \approx e^{s_-(\mathbf{k}) \eta} \approx 1\,,
\end{equation}
the solutions are still spherically symmetric. Therefore
\begin{equation}
\left\langle f^i\left(\mathbf{r},\eta\approx 0\right)
f^j\left(\mathbf{r}',\eta\approx 0\right)\right\rangle =
\int \frac{d^3 \mathbf{k}}{(2\pi)^{3/2}} 
e^{i \mathbf{k}.(\mathbf{r}-\mathbf{r}')}
\left[\delta^{ij} - \hat k^i \hat k^j\right] F(k)\,.\label{isotropiceq}
\end{equation}
Using the properties of the vectors $\bar\varepsilon_{\pm}$ and its relation with the transverse projector
\bea
\sum_{\lambda=+,-}\varepsilon^i_\lambda(\bm{ {\rm k}})\varepsilon^j_\lambda(\bm{ {\rm k}})=\delta^{ij}-\hat k^i \hat k^j\,,
\tea
we read from (\ref{isotropiceq}) that
\begin{eqnarray}
\left\langle f_+(\mathbf{q}) f_+(\mathbf{k})\right\rangle &=&
\left\langle f_-(\mathbf{q}) f_-(\mathbf{k})\right\rangle = 0 \label{corr1} \\
\left\langle f_-(\mathbf{q}) f_+(\mathbf{k})\right\rangle &=&
\left\langle f_+(\mathbf{q}) f_-(\mathbf{k})\right\rangle = 
\left(2\pi\right)^{3/2} \delta\left(\mathbf{q}+\mathbf{k}\right)F(k)\label{corr2}\,.
\end{eqnarray}

\subsubsection{Magnetic Helicity}
As briefly explained at the beginning of this subsection, the magnetic helicity is defined 
in a volume $V$ as \cite{BF84}
\begin{equation}
H_M(\eta) = \int_V dV~ \mathbf{a}\left(\mathbf{r},\eta\right)
\cdot \mathbf{b}\left(\mathbf{r},\eta\right)\label{helicitydef}
\end{equation}
where $\mathbf{\nabla}\times \mathbf{a} = \mathbf{b}$. As we are dealing with pure
vector fields we can work in the Coulomb gauge and write the vector potential $\mathbf{a}\left(\mathbf{r},\eta\right)$ as
\begin{equation}
a^i\left(\mathbf{r},\eta\right) = i\int \frac{d^3 \mathbf{k}}{(2\pi)^{3/2}} 
\frac{e^{i \mathbf{k}.\mathbf{r}}}{k}
\left[ e^{s_+(\mathbf{k}) \eta} b_+(\mathbf{k}) \varepsilon_-^i(\mathbf{k})
- e^{s_-(\mathbf{k}) \eta} b_-(\mathbf{k}) \varepsilon_+^i(\mathbf{k})\right]
\end{equation}
In this way we obtain
\begin{eqnarray}
H_M\left(\eta\right) &=& 
i \int \frac{d^3 \mathbf{k}}{(2\pi)^{3/2}} 
\int \frac{d^3 \mathbf{q}}{(2\pi)^{3/2}} 
\frac{\mathcal{W}\left(\mathbf{k}+\mathbf{q}\right)}{k} \nonumber\\
&\Big[& 
e^{s_+(\mathbf{k}) \eta} e^{s_+(\mathbf{q}) \eta}
b_+(\mathbf{k}) b_+(\mathbf{q})
\varepsilon_-^i(\mathbf{k})\varepsilon_+^i(\mathbf{q})
- e^{s_-(\mathbf{k})\eta} e^{s_+(\mathbf{q})\eta}
b_-(\mathbf{k}) b_+(\mathbf{q})
\varepsilon_+^i(\mathbf{k})\varepsilon_+^i(\mathbf{q})
\nonumber\\
&+& e^{s_+(\mathbf{k})\eta} e^{s_-(\mathbf{q})\eta}
b_+(\mathbf{k}) b_-(\mathbf{q})
\varepsilon_-^i(\mathbf{k})\varepsilon_-^i(\mathbf{q})
- e^{s_-(\mathbf{k})\eta}  e^{s_-(\mathbf{q})\eta}
b_-(\mathbf{k}) b_-(\mathbf{q})
\varepsilon_+^i(\mathbf{k}) \varepsilon_-^i(\mathbf{q})\Big]\nonumber\\ 
\label{magnetichelicity}
\end{eqnarray}
where we define the window function
\begin{equation}
\mathcal{W}\left(\mathbf{k}+\mathbf{q}\right) = \int_V d^3 \mathbf{r}
e^{i \left(\mathbf{k}+\mathbf{q}\right).\mathbf{r}}\label{windowfunction}
\end{equation}
with $V$ the volume over which the magnetic helicity is computed.

The statistically spherical symmetry we assumed at the beginning of the evolution, or equivalently the lack of `handedness' or preferred orientation, implies that the mean value of the magnetic helicity vanishes as we can check from (\ref{magnetichelicity}). In consequence we directly compute its rms deviation at equal times. We assume that the variables are Gaussian.
In view of the correlations given above, the terms that will survive are the ones with the same numbers of 'plus' and 'minus' signs. In addition the averages corresponding to the mean values vanish. Using the correlations (\ref{corr1})-(\ref{corr2}), the parity properties of the eigenvectors and the eigenvalues, we write
\begin{eqnarray}
\left\langle H_M^2\right\rangle &=& - 
\int \frac{d^3 \mathbf{k}}{(2\pi)^{3/2}} 
\int \frac{d^3 \mathbf{q}}{(2\pi)^{3/2}}  
\frac{\left\vert\mathcal{W}\left(\mathbf{k}+\mathbf{q}\right)\right\vert^2}{k^2}
F(k)F(q)
\nonumber\\
&&\left\{  - e^{2s_+(\mathbf{k})\eta} e^{2s_+(\mathbf{q})\eta}
\varepsilon_-^j(\mathbf{k}) \varepsilon_+^j(\mathbf{q})
\varepsilon_-^i(\mathbf{k})\varepsilon_+^i(\mathbf{q})
- e^{2s_+(\mathbf{q})\eta} e^{2s_+(\mathbf{k})\eta}
\varepsilon_-^j(\mathbf{q}) \varepsilon_+^j(\mathbf{k})
\varepsilon_-^i(\mathbf{k})\varepsilon_+^i(\mathbf{q})
\rangle \right.\nonumber\\
&& + e^{2s_+(\mathbf{k})\eta} e^{2s_-(\mathbf{q})\eta}
\varepsilon_+^j(\mathbf{k})\varepsilon_+^j(\mathbf{q})
\varepsilon_-^i(\mathbf{q})\varepsilon_-^i(\mathbf{k}) 
- e^{2s_+(\mathbf{k})\eta} e^{2s_-(\mathbf{q})\eta}
\varepsilon_-^j(\mathbf{k}) \varepsilon_-^j(\mathbf{q})
\varepsilon_-^i(\mathbf{q})\varepsilon_-^i(\mathbf{k}) 
\nonumber\\
&&  - e^{2s_-(\mathbf{k})\eta} e^{2s_+(\mathbf{q})\eta}
\varepsilon_+^i(\mathbf{k})\varepsilon_+^i(\mathbf{q})
\varepsilon_+^j(\mathbf{q}) \varepsilon_+^j(\mathbf{k}) 
+ e^{2s_-(\mathbf{k})\eta} e^{2s_+(\mathbf{q})\eta}
\varepsilon_+^i(\mathbf{k})\varepsilon_+^i(\mathbf{q}) 
\varepsilon_-^j(\mathbf{q})\varepsilon_-^j(\mathbf{k})
\nonumber\\
&& \left. - e^{2s_-(\mathbf{k})\eta} e^{2s_-(\mathbf{q})\eta}
\varepsilon_+^i(\mathbf{k})\varepsilon_-^i(\mathbf{q})
\varepsilon_-^j(\mathbf{k})\varepsilon_+^j(\mathbf{q})
- e^{2s_-(\mathbf{k})\eta} e^{2s_-(\mathbf{q})\eta}
\varepsilon_+^i(\mathbf{k})\varepsilon_-^i(\mathbf{q})
\varepsilon_-^j(\mathbf{q})\varepsilon_+^j(\mathbf{k})\right\} \,.\nonumber\\
\label{Hm-a}
\end{eqnarray}
Observe that both $s_+$ and $s_-$ determine the evolution of the rms value of the magnetic helicity.
To estimate the evolution of the rms value of $H_M$ we consider a large integration volume in (\ref{helicitydef}) turning the window function (\ref{windowfunction}) into a Dirac delta. By virtue of the orthogonality properties and the closure of the basis vectors,
\begin{equation}
\sum_{\lambda=+,-}\varepsilon_\lambda^i(\mathbf{k})\varepsilon_\lambda^i(\mathbf{k})= 3 - \hat k^i \hat k_i = 2\,,
\end{equation}
the equation (\ref{Hm-a})
becomes
\begin{eqnarray}
\left\langle H_M^2\right\rangle &=& 2\int d^3 \mathbf{k}\, 
\frac{ F^2(k)}{k^2}\, e^{2\left[\sigma_+(\mathbf{k})+\sigma_-(\mathbf{k})\right]\eta}
\label{Hm-a-dir-finfin}
\end{eqnarray}
which shows that the Weibel instability also amplifies an initial rms value of the magnetic helicity. It is interesting to observe that while the magnetic field grows mainly with the largest eigenvalue $\sigma_-(\mathbf{k})$, the growth exponent of the magnetic helicity is determined by the sum of both roots  $\sigma_-(\mathbf{k})+\sigma_+(\mathbf{k})$. Indeed the rms value of the magnetic field is given by 
\begin{eqnarray}
\left\langle B^2\right\rangle &=& \int \frac{d^3\mathbf{k}}{(2\pi)^{3/2}}F(k)
\left[e^{2\sigma_+(\mathbf{k})\eta}\varepsilon_+^i(\mathbf{k})\varepsilon_+^i(\mathbf{k})
+ e^{2\sigma_-(\mathbf{k})\eta}\varepsilon_-^i(\mathbf{k})\varepsilon_-^i(\mathbf{k})\right]
\label{rmsB}\\
&\simeq & \int \frac{d^3\mathbf{k}}{(2\pi)^{3/2}}F(k)
e^{2\sigma_-(\mathbf{k})\eta}\varepsilon_-^i(\mathbf{k})\varepsilon_-^i(\mathbf{k})
\label{rmsB-b}
\end{eqnarray}
where the last expression stems from the fact that $\sigma_+(\mathbf{k})<\sigma_-(\mathbf{k})$.

We might define a coherence scale for the helical fields as the ratio of the rms 
values of the magnetic helicity to $\left\langle B^2\right\rangle$, which according to eqs. 
(\ref{Hm-a-dir-finfin}) and (\ref{rmsB-b}) will evolve as
\begin{eqnarray}
\lambda_{B_H} &\sim& \frac{\sqrt{\left\langle H_M^2\right\rangle}}{\left\langle B^2\right\rangle}
= (2\pi)^{3/2}\frac{\sqrt{2\int d^3 \mathbf{k}\, 
F^2(k)k^{-2}\, e^{2\left[\sigma_+(\mathbf{k})+\sigma_-(\mathbf{k})\right]\eta}}}{\int d^3\mathbf{k}~ F(k)
e^{2\sigma_-(\mathbf{k})\eta}\varepsilon_-^i(\mathbf{k})\varepsilon_-^i(\mathbf{k})} \label{coh-leng-1}\\
& < & 4\pi^{3/2}\frac{\int d^3 \mathbf{k}\, k^{-1}
F(k)\, e^{\left[\sigma_+(\mathbf{k})+\sigma_-(\mathbf{k})\right]\eta}}{\int d^3\mathbf{k}~ F(k)
e^{2\sigma_-(\mathbf{k})\eta}\varepsilon_-^i(\mathbf{k})\varepsilon_-^i(\mathbf{k})} \label{coh-leng-1} 
\end{eqnarray}
As $\sigma_+(\mathbf{k})<\sigma_-(\mathbf{k})$ for all modes (see eq. (\ref{eigenvalue}) and 
Fig. \ref{sMassMenos}) it is always satisfied that $\sigma_+(\mathbf{k})+\sigma_-(\mathbf{k}) < 2 \sigma_-(\mathbf{k})$ so the denominator
will grow faster than the numerator. This means that the proposed coherence length for the helical  field
will shrink as the instability evolves.

\section{Conclusions}\label{concl}

We study the feasibility of the development of a relativistic Weibel instability during the Reheating stage in the very early universe prior to the establishment of the main bulk radiation. We particularly use the second order magneto-hydrodynamic description developed in Ref. \cite{CalKan16}. We assume that during Inflation there exists an spectator charged scalar field arbitrarily coupled to gravity in its de Sitter vacuum state. Gravitational particle creation occurs at the transition from Inflation to the following stage and the field state becomes an excited many particles one which admits a hydrodynamic description.

Quantum fluctuations of the scalar field produce rms deviations of $T^{\mu\nu}$ and $J^{\mu}$. We average these quantities over super-horizon regions at the end of Inflation whose comoving size is $L_D$. The quantum rms values are translated into non-equilibrium fluid parameters through the simplest matching procedure between quantum and statistical fluctuations. In particular the energy density, related to the VEV of $T^{00}$, the anisotropic pressures, related to the rms VEV of the energy-momentum tensor, and the charge excess, related to the rms VEV of $J_0$, are the relevant ones for the unfolding of the vector instabilities.

We find that for Reheating scenarios with $N\lesssim 4$ \textit{e-folds}, which correspond to lapses of
time shorter than the relaxation time $\tau$ of the fluid, the unstable exponential growth of the magnetic
field compensates the conformal dilution or even produces a net amplification for sub-horizon scales at the end of Reheating depending on the values of the gravity coupling (see Figure \ref{results_fig}). Moreover, the instability also amplifies the magnetic helicity, while concentrating it at small scales,
thus improving the chances for an helical field to survive.

Although the instability produced by the inflationary initial quantum fluctuations is efficient only at small scales compared with the scales of astrophysical interest, it might be relevant to produce strong helical magnetic fields that work as seeds for posterior amplification processes during Reheating itself or the radiation-dominated stage.

Finally we remark the relevance of the Second Order magneto-hydrodynamics to describe the dynamics of non-ideal processes during the Very Early Universe, in this case by capturing the anisotropic pressure tensor produced by the inflationary quantum fluctuations in a consistent magneto-hydrodynamic framework.

\begin{acknowledgments}
The work of E.C. and N. M. G. was supported in part by CONICET, ANPCyT and Universidad de Buenos Aires through grant UBACYT 20020170100129BA. A. K. thanks UESC for support through project UESC/PROPP $\mathrm{N}^{\circ}$ 00220.1300.1876.
\end{acknowledgments}

\appendix

\section{Charged Scalar Field}\label{ChScField}

In this section we give a nut-shell review of charged scalar fields in 
curved space-time and compute the main functions that to characterize
the quantum gas as a real relativistic plasma: the VEV of $T^{00}_Q$, the Hadamard two-point function of the traceless part of
$T^{ij}_Q$ and of the charge density $J^0_Q$. 

A charged scalar field is mathematically described by the complex
fields $\mathbf{\Phi}(x^{\mu})$ and 
$\mathbf{\Phi}^{\dagger}(x^{\mu})$, which can be written in
terms of real fields as
\begin{eqnarray}
\mathbf{\Phi}(x^{\mu}) &=& \frac{\Phi_1 + i \Phi_2}{\sqrt{2}}
\label{sf-1}\\
\mathbf{\Phi}^{\dagger}(x^{\mu}) &=& \frac{\Phi_1 - i \Phi_2}{\sqrt{2}}
\label{sf-2}
\end{eqnarray}
where $\Phi_1$ and $\Phi_2$ commute and satisfy the Klein-Gordon equation for a
massive field arbitrarily coupled to gravity
\begin{equation}
\square \Phi - \left(\mu^2 + \xi R\right)\Phi = 0\label{KG}\,.
\end{equation}
We assume that all fields and parameters are already dimensionless.
The Energy-Momentum tensor (EMT) for each of the real scalar fields is \cite{BirrDav82,ParkToms09}
\begin{eqnarray}
T_{\mu\nu}^{(\Phi)} &=& \partial_{\mu}\Phi \partial_{\nu}\Phi 
- \frac{1}{2}g_{\mu\nu}\left[g^{\alpha\beta}
\partial_{\alpha}\Phi\partial_{\beta}\Phi 
+ \left(\mu^2 + \xi R\right)\Phi^2\right] 
+ \xi R_{\mu\nu}\Phi^2 
- \xi\left[\nabla_{\mu}\partial_{\nu} - g_{\mu\nu}
\square \right] \Phi^2
\,.\nonumber\\
\label{sf-emt}
\end{eqnarray}
The symbol $\square$ represents the relativistic D'Alambertian,
$R_{\mu\nu}$ the Ricci tensor and $R$ the scalar
curvature. The dimensionless mass of the field is $\mu=m/H$ and $\xi$ is the coupling to the curvature. The values $\xi=0$ and $\xi =1/6$ correspond to the minimal and conformal coupling respectively. The operator $\square$ applied to $\Phi^2$ can be written using the Klein-Gordon equation as
\begin{equation}
\square \Phi^2 = 2\partial_{\alpha}\Phi
\partial^{\alpha}\Phi + 2\mu^2\Phi^2 + 2\xi R \Phi^2\label{sf2-dal}\,.
\end{equation}
The EMT for each real field is then given by eq. (\ref{emt-KG-body}), with the trace
\begin{eqnarray}
T^{(\Phi)\mu}_{\mu} &=& -\partial^{\alpha}\Phi\partial_{\alpha}\Phi
-2\mu^2\Phi^2 - \xi R \Phi^2 + 3\xi\square\Phi^2\label{trace-1}\\
&=&(6\xi -1)\left[ \partial^{\alpha}\Phi\partial_{\alpha}\Phi
- \xi  R\Phi^2\right] + (6\xi - 2)\mu^2\Phi^2\label{trace-2}\,.
\end{eqnarray}
Rescaling the scalar field as
\begin{equation}
\Phi = \frac{\varPhi}{a} \label{conf-sf}\,,
\end{equation}
expr. (\ref{emt-KG-body}) becomes $T_{\mu\nu}=a^{-2}\tilde T_{\mu\nu}$ with $\tilde T_{\mu\nu}$
given by eq. (\ref{conf-KG-body})
The electric four current of a charged scalar field is defined as
\cite{ItzZub80}
\begin{equation}
J_{\mu}\left(x^{\nu}\right) = 
ie \left[\mathbf{\Phi}^{\dagger}\left(x^{\nu}\right)
\partial_{\mu}\mathbf{\Phi}\left(x^{\nu}\right)
- \partial_{\mu}(\mathbf{\Phi}^{\dagger}\left(x^{\nu}\right))
\mathbf{\Phi} \left(x^{\nu}\right)\right]\label{el-curr}
\end{equation}
which in terms of the real fields is
\begin{equation}
J_{\mu}\left(x^{\nu}\right) = - e \left[\Phi_1 \left(x^{\nu}\right)
\partial_{\mu}\Phi_2 \left(x^{\nu}\right) - \Phi_2\left(x^{\nu}\right)
\partial_{\mu}\Phi_1\left(x^{\nu}\right) \right]\label{ell-curr-2}
\end{equation}
After replacing (\ref{conf-sf}) for each field we have
that $J_{\mu}=a^{-2}\tilde J_{\mu}$ with $\tilde J_{\mu}$
given in eq. (\ref{ell-curr-3-body})

\subsection{Scalar Field in de Sitter Space-time}

We now present the solutions of eq. (\ref{KG}) for Inflation, 
where the scale factor is $a = (1-\eta)^{-1}$. We consider that $\eta = 0$ and $a_I(0) = a'_I(0) = 1$ at the end  of Inflation. In terms of the quantum creation and
annihilation operators the scalar field $\varPhi$ reads
\begin{equation}
\varPhi(\hat r, \eta) = \int \frac{d^3 k}{(2\pi)^{3/2}}
\left[e^{i\hat k\cdot \hat r}\varphi_k(\eta) a_k +
e^{-i\hat k\cdot \hat r}\varphi^{*}_k (\eta) a^{\dagger}_k\right]
\label{phi-fourier}
\end{equation}
where $\varphi_k$ and $\varphi^*_k$ are the solutions of the 
Fourier transformed, dimensionless field equation
\begin{equation}
\varphi^{\prime\prime}_k + \left[ k^2 + \mu^2a^2
+ \left(\xi - \frac{1}{6}\right)\frac{a^{\prime\prime}}{a}\right]
\varphi_k = 0 \label{phi-eq}
\end{equation}
where again $\mu = m/H$. The mode equation
(\ref{phi-eq}) becomes a Bessel equation for $f_k$ after the change $\varphi_k = a^{-1/2}(\eta)f_k(1-\eta)$, namely
\begin{equation}
f_k^{\prime\prime} + \frac{f^{\prime}_k}{(1-\eta)}
+ \left[ k^2 + \frac{(\mu^2 + 12\xi - 9/4)}{(1-\eta)^2}\right]f_k = 0
\end{equation}
The normalized, positive frequency modes for $\eta \to -\infty$ 
are
\begin{equation}
\varphi_k(\eta) = \frac{\pi^{1/2}}{2}e^{ik}
(1-\eta)^{1/2}
H_{\nu}^{(1)}\left[k(1-\eta)\right]; ~~~~
\nu = \frac{3}{2}\left(1-\frac{4}{9}\mu^2
- \frac{16}{3}\xi\right)^{1/2} \label{sf-modes-1}
\end{equation}
where $H_{\nu}^{(1)}(z)$ are the Hankel functions and the phase $e^{ik}$ is included in order to correctly match the 
$\mu=0$ limit.

To compute the vev's and the kernels that describe the statistical properties of the plasma 
we consider that during Inflation the
scalar field is in its adiabatic vacuum, which means that after the
transition to the subsequent epoch, observers in the local vacuum state
will detect a bath of particles built up from super-horizon modes.
Mathematically this means that the momenta that will build the different
functions are those with $k \ll 1$.
Therefore we can retain only the leading term in the expansion of the 
Hankel functions, the resulting expression for the modes is
\begin{equation}
\varphi_k \simeq -\frac{\pi^{1/2}}{2} 
\frac{i e^{ik}}{\sin(\nu\pi)}
\frac{1}{\Gamma(1-\nu)} \frac{2^{\nu}}{k^{\nu}}
\label{mode-asympt}\,.
\end{equation}


\begin{thebibliography}{99}

\bibitem{fermilab-rep20} R. Allahverdi, M. A. Amin, et al., \textit{The first three seconds: a review of possible expansion
	histories of the Early Universe}, Fermilab-Pub-20-242-A, KCL-PH-TH/2020-33, arXiv:2006.16182 [astro-ph.CO] (2020).

\bibitem{donough-20} E. McDonough, \textit{The cosmological heavy ion collider: Fast thermalization after cosmic
	inflation}, Phys. Lett. B \textbf{809}, 135755 (2020)


\bibitem{CKPWS-18} M. P. DeCross, D. I. Kaiser, A. Prabhu, C. Prescod-Weinstein and E. I. Sfakianakis,
\textit{Preheating after multifluid inflation with nonminimal couplings. I. Covariant formalism and attractor behavior},
Phys. Rev. D \textbf{97}, 023526 (2018); \textit{Preheating after multifield inflation with nonminimal couplings. II. Resonance structure}, Phys. Rev. D \textbf{97}, 023527 (2018).

\bibitem{AlSteTuWil82} A. Albrecht, P. J. Steinhardt, M. S. Turner and F. Wilczek, \textit{Reheating an
	Inflationary Universe}, Phys. Rev. Lett. \textbf{48}, 1437 (1982).

\bibitem{KoLiSta97} L. Kofman, A. Linde and A. A. Starobinsky, \textit{Towards the theory of reheating 
	after inflation}, Phys. Rev. D \textbf{56}, 3258 (1997).

\bibitem{BraNaRa} R. Brandenberger, R. Namba and R. O. Ramos, \textit{Kinetic equilibration after preheating}, arXiv:1908.09866 [hep-ph] (2019).

\bibitem{Loz19} K. Lozanov, \textit{Lectures on Reheating After Inflation},
2018 MPA Lecture Series on Cosmology, arXiv:1907.04402 [astro-ph.CO] (2019).

\bibitem{MirtNozAz21} F. S. Mirtalebian, K. Nozari and T. Azizi, \textit{Reheating after an
	Inflationary Universe from a Perfect Fluid and Its Comparison with Observational Data}, 
Astrophys. J \textbf{907}, 107 (2021).


\bibitem{calkanma98} E. Calzetta, A. Kandus and F. D. Mazzitelli, \textit{Primordial magnetic fields induced by cosmological particle creation}, Phys. Rev. D \textbf{57}, 7139 (1998).

\bibitem{FiGru01} F. Finelli and A. Gruppuso, \textit{Resonant 
	amplification of gauge fields in expanding universe},
Phys. Lett. B \textbf{502}, 216 (2001).

\bibitem{KKT11} A. Kandus, K. Kunze and C. Tsagas, \textit{Primordial
	Magnetogenesis}, Phys. Rept. \textbf{505},1 (2011).

\bibitem{Ko14} T. Kobayashi, \textit{Primordial Magnetic Fields from the
	Post-Inflationary Universe}, JCAP \textbf{1405}, 040 (2014).

\bibitem{vach20} T. Vachaspati, \textit{Progress on Cosmological Magnetic Fields}, arXiv:2010.10525 (2020).

\bibitem{CalHu08} E. Calzetta and B. L. Hu, \textit{Nonequilibrium
	Quantum Field Theory}, Cambridge Univ. Press, Cambridge, UK
(2008).

\bibitem{Rom-Rom-19} P. Romatschke and U. Romatschke, \textit{Relativistic Fluid Dynamics In and Out of Equilibrium And 
	Applications to Relativistic Nuclear Collisions}, Cambridge Univ. Press, Cambridge, UK (2019).

\bibitem{GraCal} M. Graña and E. Calzetta, \textit{Reheating and Turbulence}, Phys. Rev. D \textbf{65}, 063522 (2002)

\bibitem{CalKan10} E. Calzetta and A. Kandus \textit{Primordial Magnetic Field Amplification from
	Turbulent Reheating}, JCAP \textbf{08}, 007 (2010).

\bibitem{tanos} L. Rezzolla and O. Zanotti, \textit{Relativistic Hydrodynamics} (Oxford University Press,
Oxford, 2013).

\bibitem{Eck40} C. Eckart, \textit{The Thermodynamics of Irreversible
	Processes. III. Relativistic Theory of the Simple Fluid},
Phys. Rev. \textbf{58}, 919 (1940)

\bibitem{LL59} L. D. Landau and E. M. Lifshitz, \textit{Fluid
	Mechanics}, Pergamon Press, Oxford (1959).

\bibitem{VanBiro12} P. V\'an and T. S. Bir\'o, \textit{First order and
	stable relativistic dissipative hydrodynamics}, Phys. Lett. B
\textbf{709}, 106 (2012).

\bibitem{Nor1} F. S. Bemfica, M. Disconzi, and J. Noronha, \textit{Causality and existence of solutions of relativistic viscous 
	fluid dynamics with
	gravity}, Phys. Rev. D \textbf{98}, 104064  (2018).

\bibitem{Nor2} F. S. Bemfica, M. Disconzi, and J. Noronha, \textit{Nonlinear causality of general first-order relativistic viscous
	hydrodynamics}, Phys. Rev. D \textbf{100}, 104020  (2019).

\bibitem{Nor3} F. S. Bemfica, M. Disconzi, and J. Noronha, \textit{General-Relativistic Viscous Fluid Dynamics}, arXiv:2009.11388  (2020).

\bibitem{Kov19} P. Kovtun, \textit{First-order relativistic hydrodynamics
	is stable}, JHEP \textbf{10}, 034 (2019).

\bibitem{DFNR20} A. Das, W. Florkowski, J. Noronha and R. Ryblewski, \textit{Equivalence between first-order causal and stable hydrodynamics and Israel-Stewart theory for boost-invariant systems with a constant 
	relaxation time}, Phys. Lett. B \textbf{806}, 135525 (2020).

\bibitem{PreRuRe20} A. L. García-Preciante, M. E. Rubio and O. A. Reula, \textit{Generic instabilities in the relativistic Chapman-Enskog heat conduction law}, J. Stat. Phys. \textbf{181}, 246 (2020). 

\bibitem{HouKov20} Raphael E. Hoult, Pavel Kovtun, \textit{Stable and 
	causal relativistic Navier-Stokes equations}, JHEP \textbf{06}, 067 (2020).

\bibitem{Isr76} W. Israel, \textit{Nonstationary irreversible thermodynamics: A causal relativistic theory}, Ann. Phys. (NY) 
\textbf{100}, 310 (1976).

\bibitem{IS76} W. Israel and J. M. Stewart, \textit{Thermodynamics of
	nonstationary and transient effects in a relativistic gas}, Phys. Lett. A \textbf{58}, 213 (1976).

\bibitem{IS79a} W. Israel and M. Stewart, \textit{Transient relativistic thermodynamics and kinetic theory}, Ann. Phys. (NY) \textbf{118}, 341 
(1979).

\bibitem{IS79b} W. Israel and M. Stewart, \textit{On transient 
	relativistic thermodynamics and kinetic theory. II}, Proc. R. Soc. London, Ser. A \textbf{365}, 43 (1979).

\bibitem{IS80} W. Israel and J. M. Stewart, \textit{Progress in 
	relativistic thermodynamics and electrodynamics of continuous media}, in General Relativity and Gravitation, edited by A. Held, Plenum, New York \textbf{2}, 491 (1980).

\bibitem{BiDaHaPubRo20} R. Biswas, A. Dash, N. Haque, S. Pub and V. Roya,
\textit{Causality and stability in relativistic viscous non-resistive magneto-fluid dynamics},
JHEP \textbf{10}, 171 (2020).

\bibitem{PaDaBisRoy21a} A. K. Panda, A. Dash, R. Biswas and V. Roy \textit{Relativistic resistive
	dissipative magnetohydrodynamics from the relaxation time approximation},

\bibitem{PaDaBisRoy21b} A. K. Panda, A. Dash, R. Biswas and V. Roy \textit{Relativistic non-resistive
	viscous magnetohydrodynamics from the kinetic theory: a relaxation time approach}, Phys. Rev. D \textbf{104},
054004 (2021).


\bibitem{LMR86} I. S. Liu, I. M\"uller and T. Ruggeri, 
\textit{Relativistic thermodynamics of gases}, Ann. Phys. \textbf{169},
191 (1986).

\bibitem{GL90} R. Geroch and L. Lindblom, \textit{Dissipative relativistic
	fluid theories of divergence type}, Phys. Rev. D \textbf{41}, 1855 (1990).

\bibitem{GL91} R. Geroch and L. Lindblom, \textit{Causal theories of
	dissipative relativistic fluids},  Ann. Phys. (NY) \textbf{207},
394 (1991).

\bibitem{ReNa95} O. A. Reula and G. B. Nagy, \textit{On the causality of 
	a dilute gas as a dissipative relativistic fluid theory of divergence
	type}, J. Phys. A \textbf{28}, 6943 (1995).

\bibitem{PRCal09} J. Peralta-Ramos and E. Calzetta, \textit{Divergence-type
	nonlinear conformal hydrodynamics}, Phys. Rev. D \textbf{80}, 126002
(2009).

\bibitem{PRCal10} J. Peralta-Ramos and E. Calzetta, \textit{Divergence-type
	2+1 dissipative hydrodynamics applied to heavy-ion collisions}, Phys. Rev. C \textbf{82}, 054905 (2010).

\bibitem{PRCal13} J. Peralta-Ramos and E. Calzetta, \textit{Macroscopic approximation to relativistic kinetic theory from a nonlinear closure}, Phys. Rev. D \textbf{87}, 034003 (2013). 

\bibitem{Cal15} E. Calzetta, \textit{Hydrodynamic approach to boost
	invariant free-streaming}, Phys. Rev. D \textbf{92}, 045035 (2015).

\bibitem{Cal98} E. Calzetta, \textit{Relativistic fluctuating hydrodynamics} Class. Quant. Grav. \textbf{15}, 653 (1998).

\bibitem{LeReRu18} L. Lehner, O. A. Reula and M. E. Rubio,
\textit{A Hyperbolic Theory of Relativistic Conformal Dissipative Fluids}, Phys. Rev. D \textbf{97}, 024013 (2018).

\bibitem{CanCal20} L. Cantarutti and E. Calzetta, \textit{Dissipative-type
	theories for Bjorken and Gubser flows}, Int. J. Mod. Phys. A
\textbf{35}, 2050074 (2020).

\bibitem{PeCa21} G. Perna and E. Calzetta, \textit{Linearized dispersion relations in viscous relativistic hydrodynamics}, [arXiv:2108.01114] (2021).

\bibitem{MGKanCal20} N. Mir\'on-Granese, A. Kandus and E. Calzetta, 
\textit{Nonlinear fluctuations in relativistic causal fluids}, JHEP \textbf{07}, 064 (2020).

\bibitem{Cal20} E. Calzetta, \textit{Fully developed relativistic turbulence}, 
Phys. Rev. D \textbf{103}, 056018 (2021).

\bibitem{DeKoRi10} G.S. Denicol, T. Koide, and D.H. Rischke, 
\textit{Dissipative relativistic fluid dynamics: a new way to derive the equations of motion from kinetic theory}, Phys. Rev. Lett. \textbf{105},
162501 (2010).

\bibitem{BDKMNR11} B. Betz, G.S. Denicol, T. Koide, E. Moln\'ar, H. Niemi,
and D.H. Rischke, \textit{Second order dissipative fluid dynamics from kinetic theory}, Eur. Phys. J. Conf. \textbf{13}, 07005 (2011).

\bibitem{DNNR11} G.S. Denicol, J. Noronha, H. Niemi, and D.H. Rischke,
\textit{Origin of the relaxation time in dissipative fluid dynamics},
Phys. Rev. D \textbf{83}, 074019 (2011).

\bibitem{DMNR12} G. S. Denicol, E. Moln\'ar, H. Niemi and D. H. Rischke,
\textit{Derivation of fluid dynamics from kinetic theory with the 14 moment
	approximation}, Eur. Phys. J. A \textbf{48} 11 (2012).

\bibitem{DNMR12} G. S. Denicol, H. Niemi, E. Moln\'ar and D. H. Rischke,
\textit{Derivation of transient relativistic fluid dynamics from the Boltzmann equation} ,Phys. Rev. D \textbf{85}, 114047 (2012); 
Phys. Rev. D \textbf{91}, 039902 (2015).

\bibitem{DNBMXRG14} G. S. Denicol, H. Niemi, I. Bouras, E. Moln\'ar, 
Z. Xue, D. H. Rischke, and C. Greiner, \textit{Solving the heat-flow 
	problem with transient relativistic fluid dynamics}, Phys. Rev. D
\textbf{89}, 074005 (2014).

\bibitem{DN12} G. S. Denicol and H. Niemi, \textit{Derivation of transient
	relativistic fluid dynamics from the Boltzmann equation for a multi-component system}, Nuc. Phys. A. \textbf{904-905}, 369c (2013).

\bibitem{MNDR14} E. Moln\'ar, H. Niemi, G. S. Denicol, and D. H. Rischke,
\textit{Relative importance of second-order terms in relativistic dissipative fluid dynamics}, Phys. Rev. D \textbf{89}, 074010 (2014).

\bibitem{ND14} H. Niemi and G. S. Denicol, \textit{How large is the Knudsen
	number reached in fluid dynamical simulations of ultrarelativistic heavy 
	ion collisions?}, arXiv:1404.7327 (2014).

\bibitem{Strick14a} M. Strickland, \textit{Anisotropic Hydrodynamics: 
	Three lectures}, Act. Phys. Pol. B \textbf{45}, 2355 (2014).

\bibitem{Strick14b} M. Strickland, \textit{Anisotropic Hydrodynamics:
	Motivation and Methodology}, Nucl. Phys. A \textbf{926}, 92 (2014).

\bibitem{FMRT15} Wojciech Florkowski, Ewa Maksymiuk, Radoslaw Ryblewski, Leonardo Tinti, \textit{Anisotropic hydrodynamics for mixture of quark 
	and gluon fluids}, Phys. Rev.C \textbf{92}, 054912 (2015).

\bibitem{FRST16} Wojciech Florkowski, Radoslaw Ryblewski, Michael Strickland and Leonardo Tinti, \textit{Non-boost-invariant dissipative hydrodynamics}, Phys. Rev. C \textbf{94}, 064903 (2016).

\bibitem{NMR17} H. Niemi, E. Moln\'ar, and D. H. Rischke, 
\textit{The right choice of moment for anisotropic fluid dynamics},
Nucl. Phys. A \textbf{967}, 409 (2017).

\bibitem{BHS14} D. Bazow, U. Heinz, M. Strickland,
\textit{Second-order (2+1)-dimensional anisotropic hydrodynamics},
Phys. Rev. C \textbf{90}, 054910 (2014).

\bibitem{MNR16a} E. Moln\'ar, H. Niemi, D. H. Rischke, \textit{Derivation
	of anisotropic dissipative fluid dynamics from the Boltzmann equation},
Phys. Rev. D \textbf{93}, 114025 (2016).

\bibitem{MNR16b} E. Moln\'ar, H. Niemi, D. H. Rischke, \textit{Closing 
	the equations of motion of anisotropic fluid dynamics by a judicious 
	choice of a moment of the Boltzmann equation}, Phys. Rev. D \textbf{94},
125003 (2016).


\bibitem{ChapCow} S. Chapman and T. G. Cowling, \textit{The Mathematical Theory of 
	Non-Uniform Gases}, 3rd. ed. ,Cambridge Univ. Press, Cambridge (1970).

\bibitem{Nahuel18} N. Mir\'on-Granese and E. Calzetta, \textit{Primordial gravitational wave amplification from causal fields}, Phys. Rev. D \textbf{97}, 023517 (2018).

\bibitem{nahuel20} N. Mir\'on-Granese, \textit{Relativistic viscous effects on the primordial gravitational waves spectrum}, JCAP \textbf{2021} 008 (2021).

\bibitem{DexNoNoYuRa21} V. Dexheimer, J. Noronha, J. Noronha-Hostler, N. Yunes and C. Ratti,
\textit{Future physics perspectives on the equation of state from heavy ion collisions to
	neutron stars}, J. Phys. G: Nucl. Part. Phys. \textbf{48}, 073001 (2021).

\bibitem{KaKuTs11} A. Kandus, K. Kunze and C. G. Tsagas, \textit{Primordial Magnetogenesis}, Phys. Rept.
\textbf{505}, 1 (2011).

\bibitem{kacalmawa00} A. Kandus, E. Calzetta, F. D. Mazzitelli and C. E. M. Wagner, textit{Cosmological
	magnetic fields from gauge-mediated supersymmetry-breaking models}, Phys. Lett. B \textbf{472}, 287 (2000).

\bibitem{CalKan16} E. Calzetta and A. Kandus, \textit{A hydrodynamic approach to the study of anisotropic instabilities in 
	dissipative relativistic plasmas}, Int. J. Mod. Phys. A \textbf{31}, 1650194 (2016).

\bibitem{WiRyuScle-al12} L. M. Widrow, D. Ryu, D. R. G. Schleicher, K. Subramanian, C. G. Tsagas and 
R. A. Treumann, \textit{The First Magnetic Fields}, Space Sci. Rev. \textbf{166}, 37 (2012).

\bibitem{weibel59} E. S. Weibel, \textit{Spontaneously growing transverse waves in a plasma due to an anisotropic velocity distribution}, Phys. Rev. Lett. \textbf{2}, 83 (1959).

\bibitem{fried1959} B. D. Fried, \textit{Mechanism for instability of transverse plasma waves}, Phys. Fluids \textbf{2} 337-337 (1959).

\bibitem{shlick04} R. Schlickeiser, \textit{Covariant kinetic dispersion theory of linear waves in anisotropic plasmas. I. General dispersion relations, bi-Maxwellian distributions and nonrelativistic limits}, Phys. Plasmas \textbf{11}, 5532 (2004).

\bibitem{acht07a} A. Achterberg and J. Wiersma, \textit{The Weibel instability in relativistic plasmas - I. Linear theory}, A\&A \textbf{475}, 1 (2007).

\bibitem{acht07b} A. Achterberg, J. Wiersma and C. A. Norman, \textit{The Weibel instability in relativistic plasmas - II. Nonlinear theory and stabilization mechanism}, A\&A \textbf{475}, 19 (2007).

\bibitem{basu02} B. Basu, \textit{Moment equation description of Weibel instability}, Phys. Plasmas \textbf{9}, 5131 (2002).

\bibitem{BretDeu06} A. Bret and C. Deutsch, \textit{A fluid approach to linear beam plasma electromagnetic instabilities}, Phys. Plasmas \textbf{13}, 042106 (2006).

\bibitem{Bret06} A. Bret, \textit{A simple analytical model for the Weibel instability in the non-relativistic regime}, Phys. Lett. A \textbf{359}, 52 (2006).


\bibitem{mrowczynski2006} C. Manuel and S. Mr\'owczy\ifmmode \acute{n}\else \'{n}\fi{}ski, \textit{Chromohydrodynamic approach to the unstable quark-gluon plasma}, Phys. Rev. D \textbf{74} 105003 (2006).


\bibitem{mrowczynski2017} S. Mrówczyński, B. Schenke, and M. Strickland, \textit{Color instabilities in the quark–gluon plasma}, Physics Reports \textbf{682} 1-97 (2017).



\bibitem{ItzZub80} C. Itzykson and J. B. Zuber, \textit{Quantum Field Theory}, Dover, New York (2005).

\bibitem{BirrDav82} N. D. Birrell and P. C. W. Davies, \textit{Quantum Fields in Curved Space}, Cambridge University Press, Cambridge, UK (1982).

\bibitem{ParkToms09} L. E. Parker and D. J. Toms, \textit{Quantum Field
	Theory in Curved Spacetime}, Cambridge Univ. Press, Cambridge, UK (2009).

\bibitem{Marsh16} D. J. Marsh, \textit{Axion Cosmology}, Phys. Rept. \textbf{643}, 1 (2016).

\bibitem{BauGreWa16} D. Baumann, D. Green and B. Wallisch, \textit{New Target for Cosmic Axion Searches},
Phys. Rev. Lett. \textbf{117}, 171301 (2016)

\bibitem{HOTW17} L. Hui, J. P. Ostriker, S. Tremaine and E. Witten, \textit{Ultralight Scalars as Cosmological
	Dark Matter}, Phys. Rev. D \textbf{95}, 043541 (2017).

\bibitem{GhSa20} D. Ghosh and D. Sachdeva, \textit{Constraints on Axion-Lepton Coupling from Big Bang Nucleosynthesis},
JCAP \textbf{2020}, 060 (2020).


\bibitem{LozAmin16} K. D. Lozanov and M. A. Amin, \textit{Equation of State and Duration to Radiation
	Domination after Inflation}, Phys. Rev. Lett. \textbf{119}, 061301 (2017).


\bibitem{BF84} M. A. Berger and G. B. Field, \textit{The topological properties of magnetic
	helicity}, J. Fluid Mech. \textbf{147}, 133 (1984).

\bibitem{biskamp93} D. Biskamp, \textit{Nonlinear Magnetohydrodynamics}, Cambridge University
Press, Cambridge, UK (1993).

\bibitem{biskamp03} D. Biskamp, \textit{Magnetohydrodynamic Turbulence}, Cambridge University
Press, Cambridge, UK (2003).


\bibitem{corn-97} J. M. Cornwall, \textit{Speculations on primordial magnetic helicity}, 
Phys. Rev. D \textbf{56}, 6146 (1997).

\bibitem{vach-91} T. Vachaspati, \textit{Magnetic fields from cosmological phase transitions},
Phys. Lett. B \textbf{265}, 258 (1991).

\bibitem{JackPi-00} R. Jackiw and S-Y. Pi, \textit{Creation and evolution of magnetic helicity},
Phys. Rev. D \textbf{61}, 105015 (2000).

\bibitem{ZhaFeVach17} Y. Zhang, F. Ferrer, and T. Vachaspati, \textit{Vacuum topology and the 
	electroweak phase transition}, Phys. Rev. D \textbf{96}, 043014 (2017).

\bibitem{CapSor14} C. Caprini and L. Sorbo, \textit{Adding helicity to inflationary
	magnetogenesis}, JCAP \textbf{1410}, 056 (2014).

\bibitem{CarGuzzSor18} C. Caprini, M. C. Guzzetti and L. Sorbo, \textit{Inflationary
	magnetogenesis with added helicity: constraints from non-gaussianities}, Class. Quant.
Grav. \textbf{35}, 124003 (2018).

\bibitem{calkan14} E. Calzetta and A. Kandus, Phys. Rev. D \textbf{89}, 083012 (2014).

\bibitem{kmlk-14} T. Kahniashvili, Y. Maravin, G. Lavrelashvili, and A. Kosowsky, 
\textit{Primordial magnetic helicity constraints from WMAP nine-year data}, 
Phys. Rev. D \textbf{90}, 083004 (2014)

















\end{thebibliography}
\end{document}